\documentclass[preprint,12pt, aps,amsmath,amssymb,nofootinbib,superscriptaddress]{revtex4} 
\usepackage[utf8]{inputenc}
\usepackage[breaklinks]{hyperref}
\usepackage{epsfig} 
\usepackage{wasysym} 
\usepackage{mathrsfs} 
\usepackage{graphicx} 
\usepackage{amsfonts} 
\usepackage{amsbsy} 
\usepackage{amscd} 
\usepackage{pstricks} 
\usepackage{multirow}
\usepackage{slashed} 
\usepackage{tikz}
\usetikzlibrary{decorations.pathmorphing,decorations.markings}
\newcommand{\beq}{\begin{equation}}
\newcommand{\eeq}{\end{equation}}
\newcommand{\bea}{\begin{eqnarray}}
\newcommand{\eea}{\end{eqnarray}}
\newcommand{\beas}{\begin{eqnarray*}}
\newcommand{\eeas}{\end{eqnarray*}}
\newcommand{\bi}{\begin{itemize}}
\newcommand{\ei}{\end{itemize}}

\def\tev{\,{\ifmmode\mathrm {TeV}\else TeV\fi}}
\def\gev{\,{\ifmmode\mathrm {GeV}\else GeV\fi}}
\def\to{\rightarrow}

\allowdisplaybreaks

\begin{document}

\title{
Searching for the minimal Seesaw models at the LHC and beyond
}

\author{Arindam Das\footnote{arindam@kias.re.kr, ORCID-ID:orcid.org/0000-0002-8934-2300}}
\affiliation{School of Physics, KIAS, Seoul 130-722, Korea}
\preprint{\today}
\begin{abstract}
The existence of the tiny neutrino mass and the flavor mixing can be naturally explained by type-I Seesaw model which is probably the simplest extension of the Standard Model (SM)
using Majorana type SM gauge singlet heavy Right Handed Neutrinos (RHNs). If the RHNs are around the Electroweak(EW)-scale having sizable mixings with the SM light neutrinos, they 
can be produced at the high energy colliders such as Large Hadron Collider (LHC) and future $100$ TeV proton-proton (pp) collider through the characteristic signatures with same sign di-lepton introducing lepton number violations(LNV). 
On the other hand Seesaw models, namely inverse Seesaw,  with small LNV parameter can accommodate EW-scale pseudo-Dirac neutrinos with sizable mixings with SM light neutrinos while satisfying the neutrino oscillation
data. Due to the smallness of the LNV parameter of such models, the `smoking-gun' signature of same-sign di-lepton is suppressed where as the RHNs in the model will manifest at the LHC and future $100$ TeV pp collider dominantly 
through the Lepton number conserving (LNC) trilepton final state with Missing Transverse Energy (MET). Studying various production channels of such RHNs we give an updated upper bound on the mixing parameters 
of the light-heavy neutrinos at the 13 TeV LHC and future $100$ TeV pp collider.

\end{abstract}
\maketitle
\tableofcontents
\clearpage
\section{Introduction}
The experimental evidence of the neutrino oscillation and flavor mixings from neutrino oscillation experiments \cite{Neut1,Neut2,Neut3,Neut4,Neut5,Neut6} indicates that the SM is not enough to explain the existence of the tiny neutrino mass and flavor mixing.  After the pioneering discovery of the $d=5$ operator \cite{Weinberg:1979sa} within the SM using the SM leptons, SM Higgs doublets and $\Delta L=2 (L= \rm{Lepton~number})$ unit of LNV, it turned out that the Seesaw mechanism \cite{Seesaw0,Seesaw1,Seesaw2,Seesaw3,Seesaw4,Seesaw5,Seesaw6} could be the simplest idea to explain the small neutrino mass and flavor mixing where the SM can be extended by the SM-gauge singlet Majorana type RHNs. 
After the Electroweak (EW) symmetry breaking, the light Majorana neutrino masses are generated after the so-called type-I Seesaw mechanism.

In type-I Seesaw, we introduce
  SM gauge-singlet Majorana RHNs $N_R^{\beta}$     
  where $\beta$ is the flavor index. $N_R^{\beta}$ couples with the SM lepton doublets 
  $\ell_{L}^{\alpha}$ and the SM Higgs doublet $H$.
The relevant part of the Lagrangian is
\bea
\mathcal{L} \supset -Y_D^{\alpha\beta} \overline{\ell_L^{\alpha}}H N_R^{\beta} 
                   -\frac{1}{2} m_N^{\alpha \beta} \overline{N_R^{\alpha C}} N_R^{\beta}  + H. c. .
\label{typeI}
\eea
where $Y_D^{\alpha \beta}$ is the Yukawa coupling, $\ell_L^{\alpha}$ is the SM $SU(2)_L$ lepton doublet and $m_N$ is the Majorana mass term. 
After the spontaneous EW symmetry breaking
   by the vacuum expectation value (VEV), 
   $ H =\begin{pmatrix} \frac{v}{\sqrt{2}} \\  0 \end{pmatrix}$, 
    we obtain the Dirac mass matrix as $M_{D}= \frac{Y_D v}{\sqrt{2}}$ suppressing the indices.
Using the Dirac and Majorana mass matrices 
  we can write the neutrino mass matrix as 
\bea
M_{\nu}=\begin{pmatrix}
0&&M_{D}\\
M_{D}^{T}&&m_{N}
\end{pmatrix}
\label{typeInu1}
\eea
where $M_D$ is the Dirac mass matrix and $m_N$ is the Majorana mass term.
Diagonalizing $M_{\nu}$ we obtain the Seesaw formula
 for the light Majorana neutrinos as 
\bea
m_{\nu} \simeq - M_{D} m_{N}^{-1} M_{D}^{T}.
\label{SeesawI}
\eea
For $m_{N}\sim 100$ GeV, we may find $Y_{D} \sim 10^{-6}$  with $m_{\nu}\sim 0.1$ eV.
However, in the general parameterization for the Seesaw formula \cite{Casas:2001sr},  
  $Y_{D}$ can be large and sizable $(\mathcal{O}(1))$, and this is the case we consider in this paper.

The searches of the Majorana RHNs can be performed by the `smoking-gun' of the same-sign di-lepton plus di-jet signal which is suppressed by the square of the light-heavy neutrino mixing parameter $|V_{\ell N}|^2\simeq |m_D m_N^{-1}|^2$. 
A comprehensive general study on the parameters of $|V_{\ell N}|^{2}$ has been given in \cite{Das:2017nvm} using the data from the neutrino oscillation experiments \cite{Neut1,Neut2,Neut3,Neut4,Neut5,Neut6}, bounds from the Lepton Flavor Violation (LFV) experiments \cite{Adam:2011ch, Aubert:2009ag, OLeary:2010hau},
Large Electron-Positron (LEP) and Electroweak Precision test \cite{deBlas:2013gla, Adriani:1992pq, Acciarri:1999qj, Achard:2001qv,
Abreu:1996pa, Blondel:2014bra, delAguila:2008pw, Akhmedov:2013hec} experiments using the non-unitarity effects \cite{Antusch:2006vwa, Abada:2007ux} applying the Casas- Ibarra conjecture \cite{Casas:2001sr, Ibarra:2010xw, Ibarra:2011xn, Dinh:2012bp, Penedo:2017knr}.
At this point we mention that  \cite{Das:2017nvm} has a good agreement with a previous analysis \cite{Rasmussen:2016njh} on the sterile neutrinos. The bounds in \cite{Das:2017nvm} has been compared with the existing results in \cite{ Achard:2001qv,delAguila:2008pw, Akhmedov:2013hec, BhupalDev:2012zg, Aad:2015xaa, CMS8:2016olu} considering the degenerate Majorana RHNs.
In case of Seesaw mechanism the Dirac Yukawa matrix $(Y_D)$ can carry the flavors where the RHN mass matrix is considered to be diagonal. This case is favored by the neutrino oscillation data as studied in \cite{Das:2017nvm, Kersten:2007vk, Xing:2009in, He:2009ua, Ibarra:2010xw, Deppisch:2010fr,Dev:2013oxa}.  Such a scenario for the Seesaw scenario is called the Flavor Non-Diagonal (FND). The other possibility of considering both of the diagonal $Y_D$ and $m_N$ is not supported by the neutrino oscillation data.

Since any number of singlets can be added in a gauge theory without contributing to anomalies, one can utilize such freedom to find a natural alternative of the low-scale realization of the Seesaw mechanism. Simplest among such scenarios is commonly known as the inverse Seesaw mechanism \cite{Mohapatra:1986aw, Mohapatra:1986bd} where a small Majorana neutrino mass originates from tiny LNV parameters rather than being suppressed by the RHN mass as done in the case of conventional Seesaw mechanism. In the inverse Seesaw model two sets of SM-singlet RHNs are introduced which are pseudo-Dirac by nature and their Dirac Yukawa couplings can be even order one, while reproducing the neutrino oscillation data. Therefore at the high energy colliders such pseudo-Dirac neutrinos can be produced through a sizable mixing with the SM light neutrinos \cite{Das:2012ze,Das:2014jxa, Das:2015toa, Das:2016akd,Das:2016hof,Das:2017pvt,Das:2017zjc,Das:2017rsu}.

In the inverse Seesaw mechanism the relevant part of the Lagrangian is given by
\bea
\mathcal{L} \supset - Y_D^{\alpha\beta} \overline{\ell_L^{\alpha}} H N_R^{\beta}- m_N^{\alpha \beta} \overline{S_L^{\alpha}} N_R^{\beta} -\frac{1}{2} \mu_{\alpha \beta} \overline{S_L^{\alpha}}S_L^{\beta^{C}} + H. c. ,
\label{InvYuk}
\eea 
where  $N_R^{\alpha}$ and $S_L^{\beta}$ are two SM-singlet heavy neutrinos
   with the same lepton numbers, $m_N$ is the Dirac mass matrix, and
   $\mu$ is a small Majorana mass matrix violating the lepton numbers.
After the EW symmetry breaking we obtain the neutrino mass matrix as 
\bea
M_{\nu}=\begin{pmatrix}
0&&M_{D}&&0\\
M_{D}^{T}&&0&&m_{N}^{T}\\
0&&m_{N}&&\mu
\end{pmatrix}.
\label{InvMat}
\eea
Diagonalizing this mass matrix we obtain the light neutrino mass matrix 
\bea
M_{\nu} \simeq M_{D} m_{N}^{-1}\mu (m_{N}^{-1})^{T} M_{D}^{T}.
\label{numass}
\eea
Note that the smallness of the light neutrino mass originates 
  from the small LNV term $\mu$. 
The smallness of $\mu$ allows the $m_{D}m_{N}^{-1}$ parameter
  could be on the order one even for an EW scale RHNs. In the inverse Seesaw
  case we can consider $Y_D$ as non-diagonal when $\mu$ and $m_N$ are diagonal
  which is called the Flavor Non-Diagonal (FND) case.
 On the other hand we can also consider the diagonal $Y_D$, $m_N$ when $\mu$ will be 
 non-diagonal. This situation is called the Flavor Democratic scenario. For the inverse Seesaw
both of the FND and FD cases are supported by the neutrino oscillation data. In this article 
we will consider the FND case from the Seesaw and the FD case from the inverse Seesaw mechanisms.

Through the Seesaw mechanism, a flavor eigenstate ($\nu$) of 
  the SM neutrino is expressed in terms of the mass eigenstates 
  of the light ($\nu_m$) and heavy ($N_m$) Majorana neutrinos such as 
\bea 
  \nu \simeq  \nu_m  + V_{\ell N} N_m,  
\eea

Using the mass eigenstates, the charged current (CC) interaction for the heavy neutrino 
  is given by 
\bea 
\mathcal{L}_{CC} \supset 
 -\frac{g}{\sqrt{2}} W_{\mu}
  \bar{e} \gamma^{\mu} P_L   V_{\ell N} N_m  + H.c., 
\label{CC}
\eea
where $e$ denotes the three generations of the charged leptons in the vector form, and 
  $P_L =\frac{1}{2} (1- \gamma_5)$ is the projection operator. 
Similarly, the neutral current (NC) interaction is given by 
\bea 
\mathcal{L}_{NC} \supset 
 -\frac{g}{2 c_w}  Z_{\mu} 
\left[ 
  \overline{N_m} \gamma^{\mu} P_L  |V_{\ell N}|^2 N_m 
+ \left\{ 
  \overline{\nu_m} \gamma^{\mu} P_L V_{\ell N}  N_m 
  + H.c. \right\} 
\right] , 
\label{NC}
\eea
 where $c_w=\cos \theta_w$ with $\theta_w$ being the weak mixing angle and $W_{\mu}, Z_{\mu}$ are the SM gauge bosons. 
 
\begin{figure*}
\begin{center}
\includegraphics[scale=0.45]{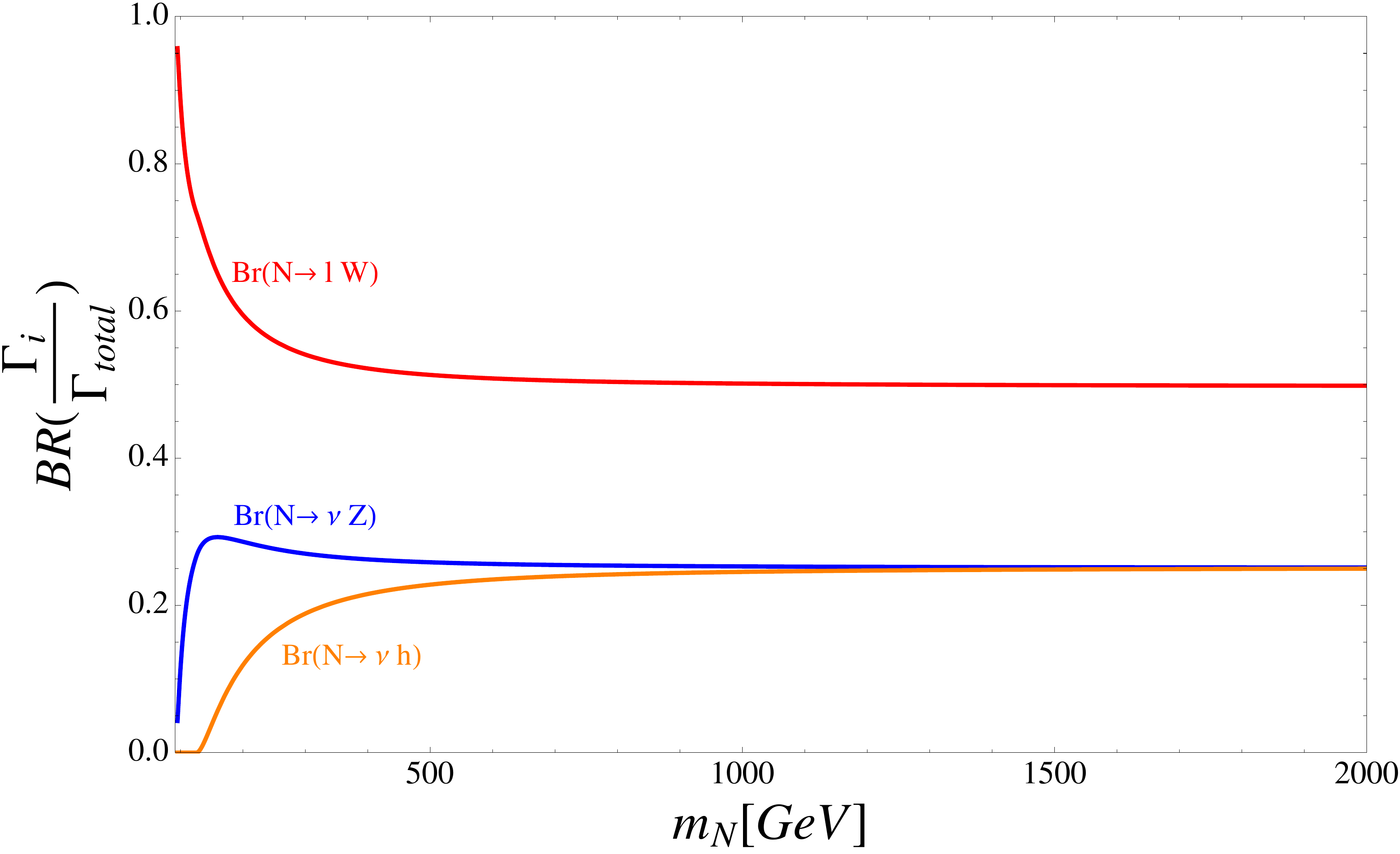}
\caption{Heavy neutrino branching ratios ($BR_i$) for different decay modes are shown with respect to its mass $\left(m_{N}\right)$ \cite{Das:2016hof}.}
\label{fig:BR}
\end{center}
\end{figure*}

 The main decay modes of the heavy neutrino are 
 $N \to \ell W$, $\nu_{\ell} Z$, $\nu_{\ell} h$. 
The corresponding partial decay widths \cite{Das:2012ze} are respectively given by
\bea
\Gamma(N \rightarrow \ell W) 
 &=& \frac{g^2 |V_{\ell N}|^{2}}{64 \pi} 
 \frac{ (m_{N}^2 - m_W^2)^2 (m_{N}^2+2 m_W^2)}{m_{N}^3 m_W^2} ,
\nonumber \\
\Gamma(N \rightarrow \nu_\ell Z) 
 &=& \frac{g^2 |V_{\ell N}|^{2}}{128 \pi c_w^2} 
 \frac{ (m_{N}^2 - m_Z^2)^2 (m_{N}^2+2 m_Z^2)}{m_{N}^3 m_Z^2} ,
\nonumber \\
\Gamma(N \rightarrow \nu_\ell h) 
 &=& \frac{ |V_{\ell N}|^2 (m_{N}^2-m_h^2)^2}{32 \pi m_{N}} 
 \left( \frac{1}{v }\right)^2.
\label{widths}
\eea 
The decay width of heavy neutrino into charged gauge bosons being twice as large as neutral one owing to the two degrees of freedom $(W^{\pm})$.
We plot the branching ratios $BR_i \left(= {\Gamma_{i}}/{\Gamma_{\rm total}}\right)$ of the respective decay modes $\left(\Gamma_{i}\right)$ with respect to the total decay width $\left(\Gamma_{\rm total}\right)$ of the heavy neutrino into $W$, $Z$ and Higgs bosons in Fig.~\ref{fig:BR} as a function of the heavy neutrino mass $\left(m_{N}\right)$.
Note that for larger values of $m_{N}$ such as $m_N \geq 1500$ GeV, the branching ratios can be obtained as 
\bea
BR\left(N\rightarrow \ell W\right) : BR\left(N\rightarrow \nu Z\right) : BR\left(N\rightarrow \nu h\right) \simeq 2: 1: 1.
\eea

In this paper we study the RHN production from various initial states such as the quark-quark $(qq)$, quark-gluon$(qg)$ and gluon-gluon$(gg)$ at the 13 TeV LHC and future $100$ TeV pp collider. We consider the photon mediated processes as well as the gluon-gluon fusion (ggF), photon-proton interactions and Vector Boson Fusion (VBF) processes to produce the RHNs. For different RHN masses $(m_{N})$, we calculate the cross section of each of the processes normalized by the square of the light-heavy neutrino mixing angles.

This paper is is organized as follows. In Sec.~\ref{prod} we calculate the production cross sections at the 13 TeV LHC and future $100$ TeV pp collider \cite{Arkani-Hamed:2015vfh} with a variety of initial states. In Sec.~\ref{decay} we study the multilepton decay modes of the RHNs at the $13$ TeV LHC and future $100$ TeV pp collider. For both of the cases we take the luminosity as $3000$ fb$^{-1}$. In Sec.~\ref{bounds} we put the bounds on the mixing angle as function of the RHN mass. Sec.~\ref{concl} is dedicated for conclusions.

\section{Production cross sections }
\label{prod}
The RHNs can be produced at the 13 TeV LHC and future $100$ TeV pp collider from various initial states. 
We consider the combined production of the heavy neutrinos from various initial states including quark-(anti)quark $(q\bar q^{\prime})$,
quark-gluon $(qg)$ and gluon-gluon $(gg)$ interactions. We also study contributions coming from the gluon-gluon fusion (ggF), 
photon-proton interactions and Vector Boson Fusion (VBF) processes to produce the RHNs at the LHC and beyond.
In this section we consider the RHN productions in association with SM leptons and jets.  To obtain the cross sections and generate the events we implement our
model in {\tt MadGraph}\cite{Alwall:2011uj} bundled with {\tt PYTHIA}\cite{Sjostrand:2006za} and {\tt DELPHES}\cite{deFavereau:2013fsa}.
The production modes of the heavy neutrinos in association with leptons and jets are proportional to the square of the mixing angle, $|V_{\ell N}|^2$.
In this section we use $|V_{\ell N}|^{2}=10^{-4}$ to estimate the cross sections \footnote{The EWPD bounds $|V_{\ell N}|^2$ have been discussed in \cite{deBlas:2013gla, delAguila:2008pw, Akhmedov:2013hec}. The universal bound has been considered to be $9\times 10^{-4}$ which rules out the possibility of the mixing angles above this value. Therefore we used a value of $|V_{\ell N}|^2 < 9\times 10^{-4}$ to calculate the cross sections.}.
\begin{itemize}
\item [(1)]  Charged current interaction mediated by $W$:\\

{\tt > generate $p p \to N \ell$}\\
{\tt > add process $p p \to N \ell j$}\\
{\tt > add process $ p p \to N \ell j j $}\\
along with the charge conjugates where $\tt j$ stands for the jets.

 In this process the the $s$-channel quark anti-quark pair of different flavors $(q\overline{q^{\prime}})$ will interact through the $W$ exchange and finally follow Eq.~\ref{CC} to produce the RHN $(N)$ in association with a lepton $(\ell)$. The corresponding Feynman diagram for $q\overline{q^{\prime}} \to N\ell$  is given in Fig.~\ref{fig:1}.

The contributions from the quark-quark interaction $(q\overline{q^{\prime}})$ to one-jet  $(N\ell j)$ are given in Fig.~\ref{fig:2} for the $t$ -channel process and
additional contributions to the one-jet process from the quark-gluon $(qg)$ interaction as shown in Fig.~\ref{fig:2a} form the $s,~t$-channel processes. In both of the cases
the $W$ is radiated form the quarks/ anti-quarks and hence follow the CC to to produce $N$ in association with $\ell$.

two-jet contributions $(N\ell jj)$ coming from the $q\overline{q^{\prime}}$ process are shown in Fig.~\ref{fig:3}.
In this case, the $s,~t$ channel contributions between the quarks are involved to produce $N$ with $\ell$ in association with two jets following the 
CC interactions at the $N$ production vertex. 
 The quark-gluon $(qg)$ processes contributing in the two-jet case are shown in Fig.~\ref{fig:4} where 
 the $s,~t$ channel contributions between the quarks and gluons are involved to produce $N$ with $\ell$ in association with two jets following the 
CC interactions at the $N$ production vertex.  
 In addition to that gluon-gluon $(gg)$ processes contributed in $N\ell jj$ are shown in Fig.~\ref{fig:5} in the $s,~t$-channel followed by the $N$ production from the 
 CC interactions with $\ell$ in association with two jets.
\begin{figure*}
\begin{center}
\includegraphics[scale=0.5]{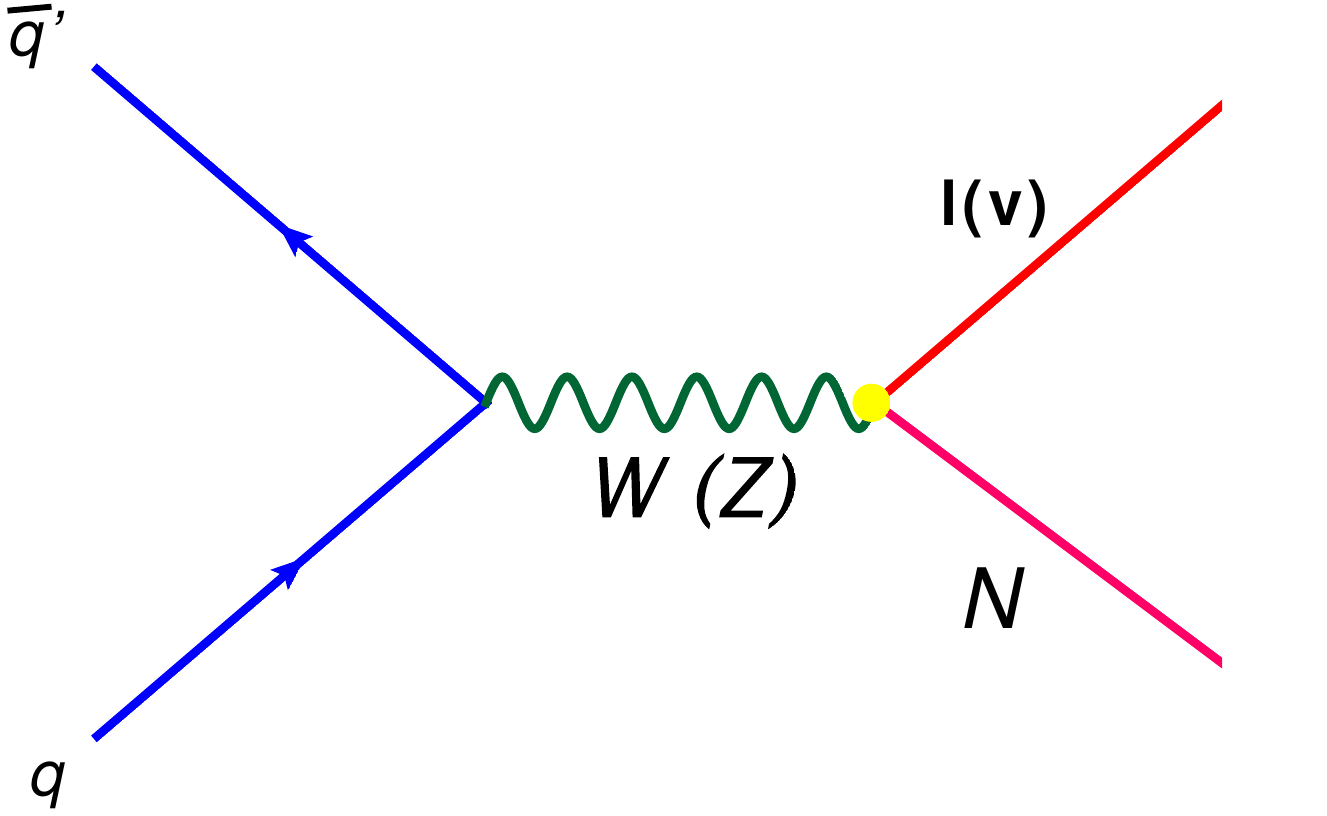}
\caption{$q\overline{q^{\prime}}(q\overline{q})$ process to produce $N\ell (N\nu)$ from the CC(NC) interaction.}
\label{fig:1}
\end{center}
\end{figure*}
\begin{figure*}
\begin{center}
\includegraphics[scale=0.7]{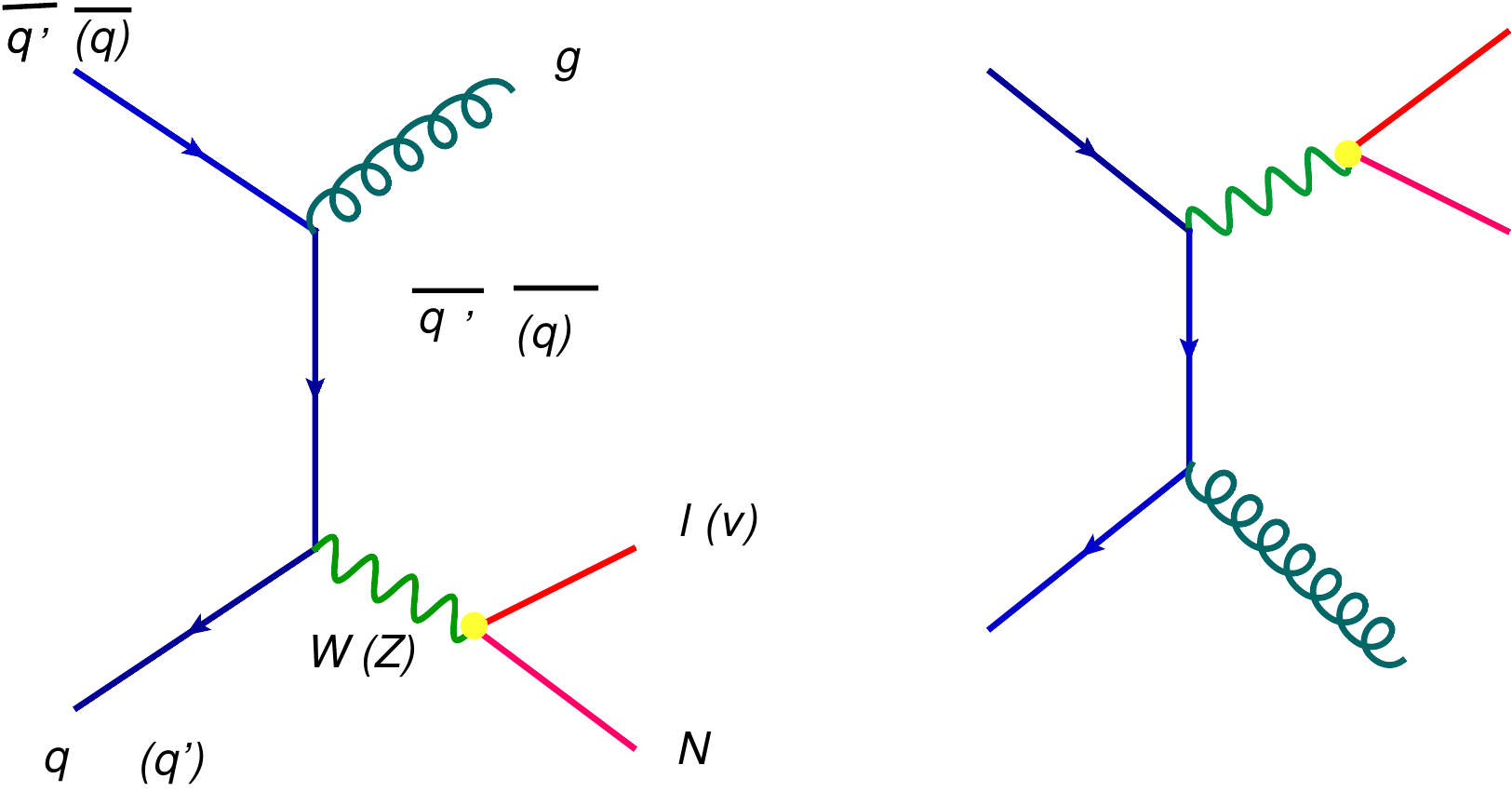}
\caption{1-jet process to produce $N$ with $\ell(\nu)$ in association with a jet from the CC(NC) interaction from $q\overline{q^{\prime}} (q\overline{q})$ initial state.}
\label{fig:2}
\end{center}
\end{figure*}
\begin{figure*}
\begin{center}\includegraphics[scale=0.7]{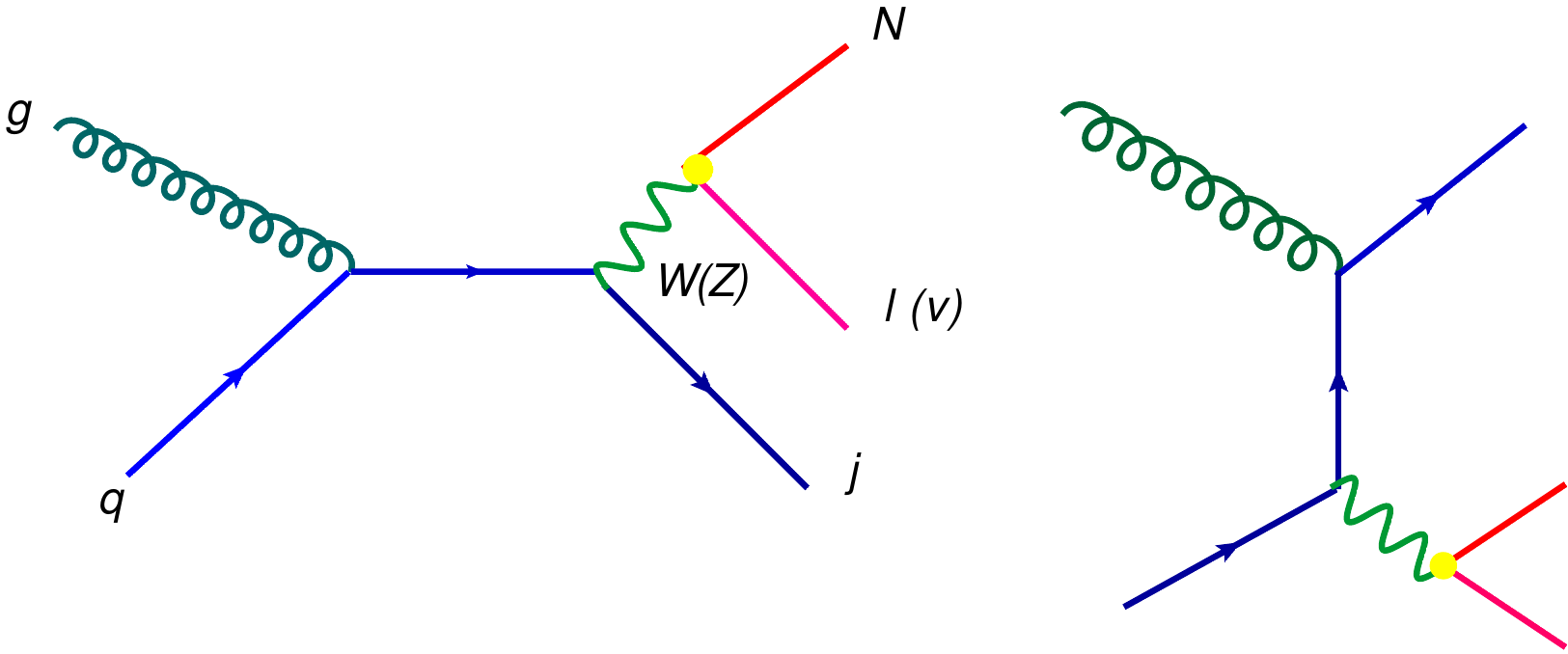}
\caption{1-jet process to produce $N$ with $\ell(\nu)$ in association with a jet from the CC(NC) interaction from $qg$ initial state.}
\label{fig:2a}
\end{center}
\end{figure*}
\begin{figure*}
\begin{center}
\includegraphics[scale=0.8]{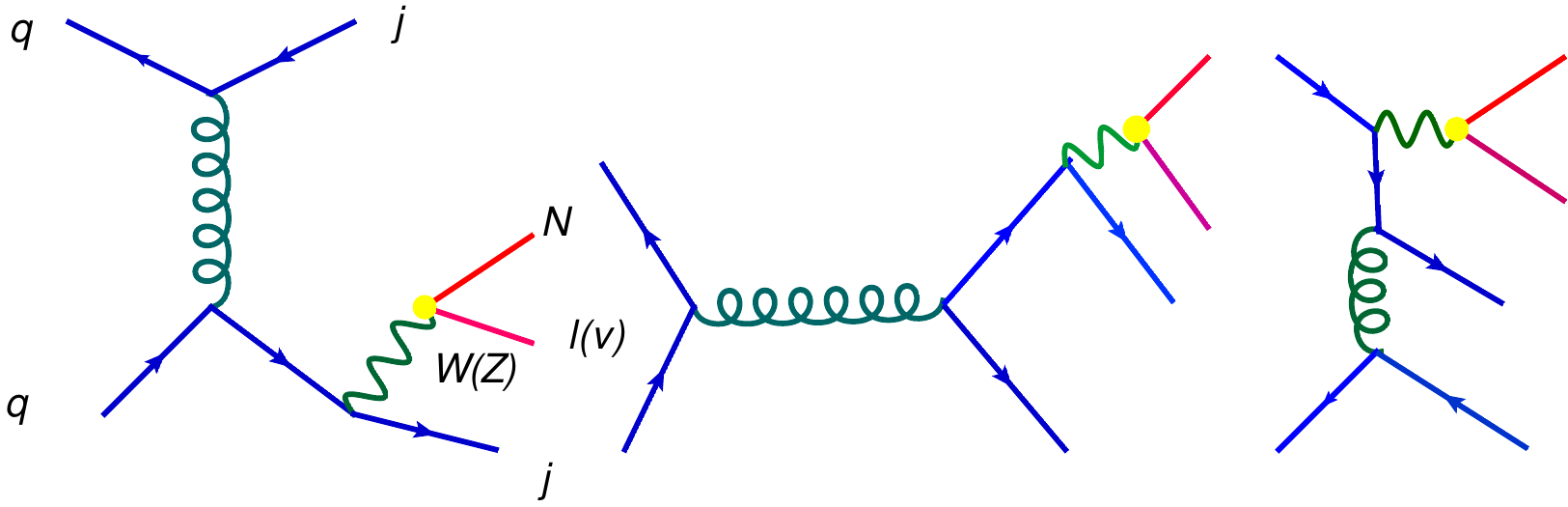}
\caption{2-jet process to produce $N$ with $\ell(\nu)$ in association with a jet from the CC(NC) interaction from $q\overline{q^{\prime}} (q\overline{q})$ initial state.}
\label{fig:3}
\end{center}
\end{figure*}
\begin{figure*}
\begin{center}
\includegraphics[scale=0.8]{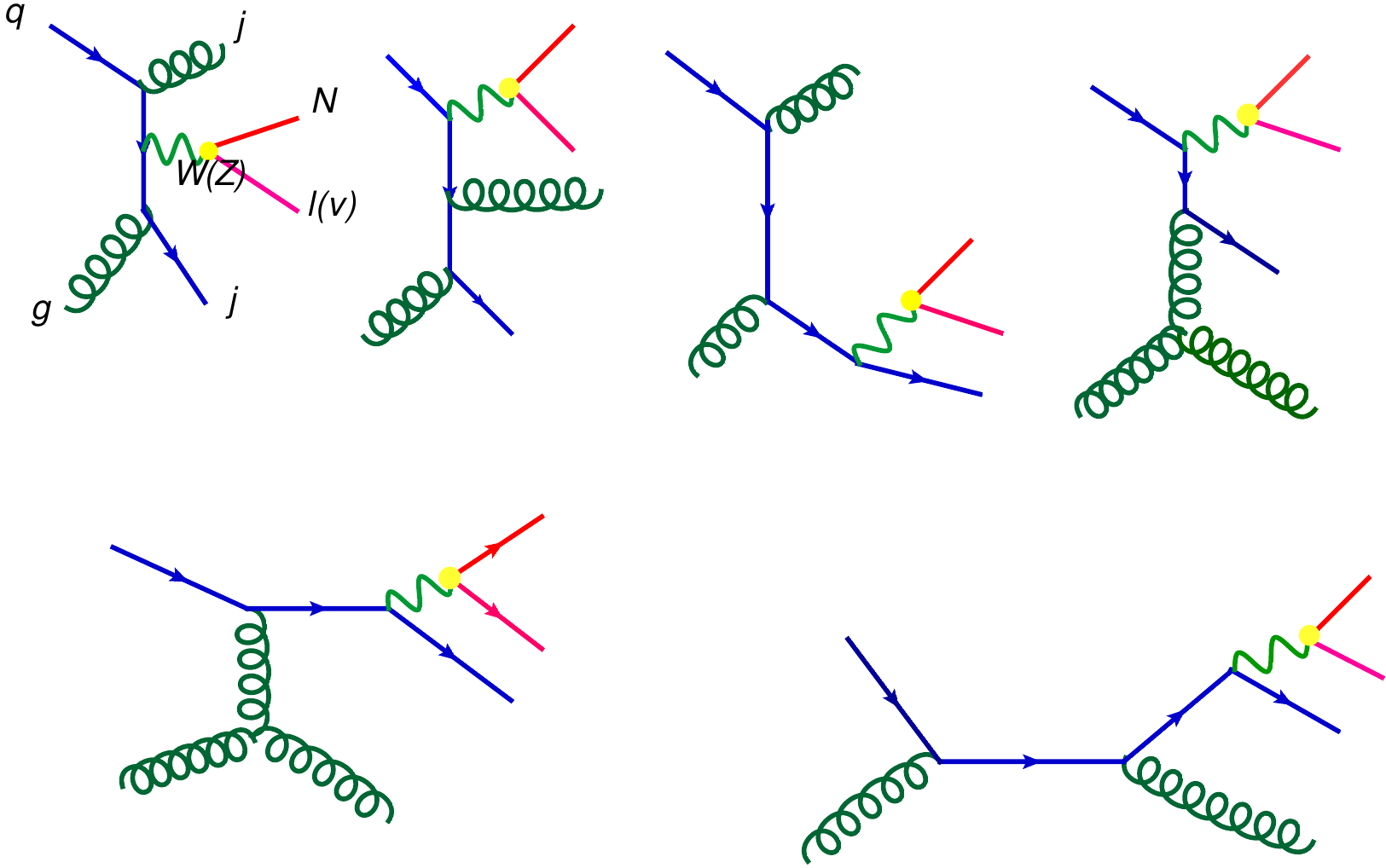}
\caption{2-jet process to produce $N$ with $\ell(\nu)$ in association with a jet from the CC(NC) interaction from $qg$ initial state.}
\label{fig:4}
\end{center}
\end{figure*}
\begin{figure*}
\begin{center}
\includegraphics[scale=0.8]{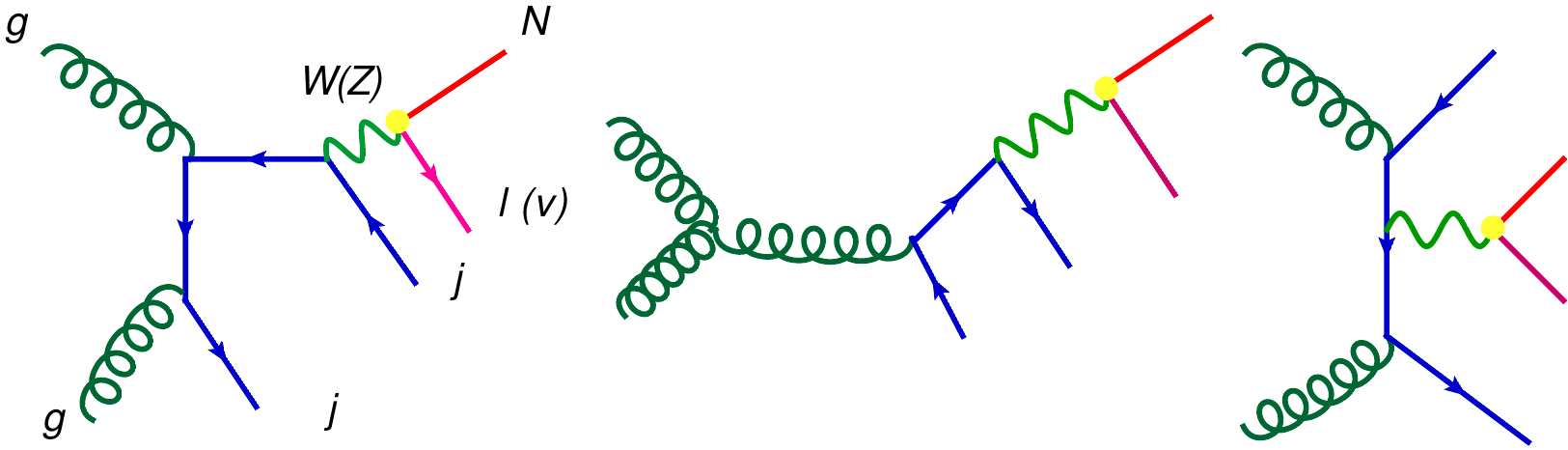}
\caption{2-jet process to produce $N$ with $\ell(\nu)$ in association with a jet from the CC(NC) interaction from $gg$ initial state.}
\label{fig:5}
\end{center}
\end{figure*}

We combine all these processes to compute the production cross section of the RHNs through the charged current interaction for 
$N\ell X$ final state where $X$ stands for $n$-jet and $ n=0, 1, 2,3$. To produce such processes we use the following
trigger cuts  
\begin{itemize}
\item [(a)]  transverse momentum of the jets, $p_{T}^{j} > 30$ GeV,
\item [(b)]  transverse momentum of the leptons, $p_{T}^{\ell} > 10$ GeV,
\item [(c)]  pseudo-rapidity of the jets, $|\eta^{j}| < 2.5$,
\item [(d)]  pseudo-rapidity of the leptons, $|\eta^{\ell}| < 2.5$
\end{itemize}
with the  Multi-Leg Matching (MLM) scheme. In the MLM matching scheme one can obtain $n$-parton events by MadEvent generator under the $k_T$ clustering algorithm. After the rewight, each final state parton can find the node where it was generated. This was $n$-parton shower can be generated $(\mu_g^2)$ in the directions of the primary partons so that the initial condition for each parton shower is kept at $\mu_g^2$. Hence running the $k_T$ clustering algorithm one can find all the jets at the resolution scale $(\tau_r)$. If all the jets match with the primary partons then the event is accepted otherwise rejected. For elaborate discussions on this mechanism for the event generators, see \cite{Matching, Matching1,Matching2, Matching3,Matching31, Matching32, Matching33, Matching34, Mangano:2002ea, Mangano:2006rw, Cooper:2011gk, Alwall:2007fs, Catani:2001cc}. In MadGraph the switch ${\tt ickkw=1}$ stands for the {\tt MLM} scheme in {\tt MadGraph} \cite{Alwall:2011uj} 
using 
\begin{itemize}
\item[{(i)}] ${\tt ickkw} =1$
\item[{(ii)}] ${\tt xqcut}= p_{T}^{j}$
\item[{(iii)}] ${\tt QCUT= Max\{xqcut+5 GeV, 1.2\ast xqcut\}}$
\end{itemize}
using  ${\tt MSTP(81)=21}$. The {\tt MSTP} switches modifies the generation procedure. In this case to switch off or on the Initial State Radiation (ISR), Final State Radiation (FSR), Multiple Interactions (MI) among the beam jets and fragmentation. 
It is also used to give the `parton skeleton' of the hard process. {\tt MSTP(81)=21} is the master switch in {\tt PYTHIA 6.4} which switches on the Multi-Interactions (MI) for the new model at the time of showering and hadronization. The ${\tt MSTP=21}$ is used for the $ p_T$ ordered shower switching on {\tt SHOWER-KT} \cite{Alwall:2008qv, Matching4, Matching5, Hoche:2006ph} in {\tt PYTHIA} \cite{Sjostrand:2006za}. The trigger cuts used on the basis of the anomalous multi-lepton searches done by the CMS \cite{CMS-trilep}. The use of cone jets or $k_T$ jets is decided by whether the parameter ${\tt xqcut}$ (specifying the minimum $k_T$ jet measure between jets, i.e. gluons or quarks (except top quarks) which are connected in Feynman diagrams) in the ${\tt MadGraph}$ is set to be $0$ or not. If ${\tt xqcut=0}$, cone jets are used, while if ${\tt xqcut > 0} $, $k_T$ jet matching is assumed. In this case, transverse momentum of jet $(p_T^j)$ and separation between the jets $(\Delta R_{jj})$ should be set to zero. For most processes, the generation speed can be improved by setting ${\tt p_T^j = xqcut}$ which is done automatically if the switch ${\tt auto-ptj-mjj}$ is set to `T' at the time of event generation using ${\tt MadGraph}$ to control the transverse momentum of the jets $(p_T^j)$ and the invariant mass of the jets $(m_{jj})$. If some jets should not be restricted this way (as in single top or Vector Boson Fusion (VBF) production, where some jets are not radiated from QCD) in that case the switch ${\tt auto-pTj-mjj}$ should be set to `F' in the MagGraph at the time of event generation. ${\tt QCUT}$ is used for the matching with the $k_T$ scheme, this is case the jet measure cutoff is used by Pythia. If the value is not given, it will be set to ${\tt max(xqcut+5,xqcut*1.2)}$, where ${\tt xqcut}$ is taken from the ${\tt MadGraph}$ \cite{Matching, Matching1}. We use such cuts to keep the analysis collinear safe. The production cross sections are shown in Figs.~\ref{fig:13} and \ref{fig:100} respectively as a function of $m_N$.
\item[(2)] Neutral current interaction mediated by $Z$:\\

{\tt > generate $p p \to N \nu$}\\
{\tt > add process $p p \to N \nu j$}\\
{\tt > add process $ p p \to N \nu j j $}\\
followed by the same production procedure as we discussed in (1). Using quark anti-quark pair of same flavor $(q\overline{q})$ at the $s$-channel by Eq.~\ref{NC} $N$ can be produced in association with a light neutrino $(\nu)$ as shown in Fig.~\ref{fig:1}.
The $N\nu j$ (in Figs.~\ref{fig:2}, \ref{fig:2a}) and $N\nu jj$ (in Figs.~\ref{fig:3}, \ref{fig:4}, \ref{fig:5}) processes are same as those have been discussed in (1). The only difference is the production vertex of $N\nu$ comes from the NC from Eq.~\ref{NC}.
Combining these final states we can denote as $N\nu X$ where $X$ stands for $n$-jet with $ n=0, 1, 2, 3$.  In this case we use the following trigger cuts
\begin{itemize}
\item [(a)]  transverse momentum of the jets, $p_{T}^{j} > 30$ GeV,
\item [(b)]  pseudo-rapidity of the jets, $|\eta^{j}| < 2.5$
\end{itemize}
and the same matching scheme used in (1). Such $p_{T}^{j}$ cuts will keep the analysis collinear safe specially in (1) and (2) to calculate the cross sections using MLM matching \cite{Matching, Matching1,Matching2, Matching3,Mangano:2002ea, Mangano:2006rw, Cooper:2011gk, Alwall:2007fs, Catani:2001cc} schemes.
\begin{figure*}
\begin{center}
\includegraphics[scale=0.8]{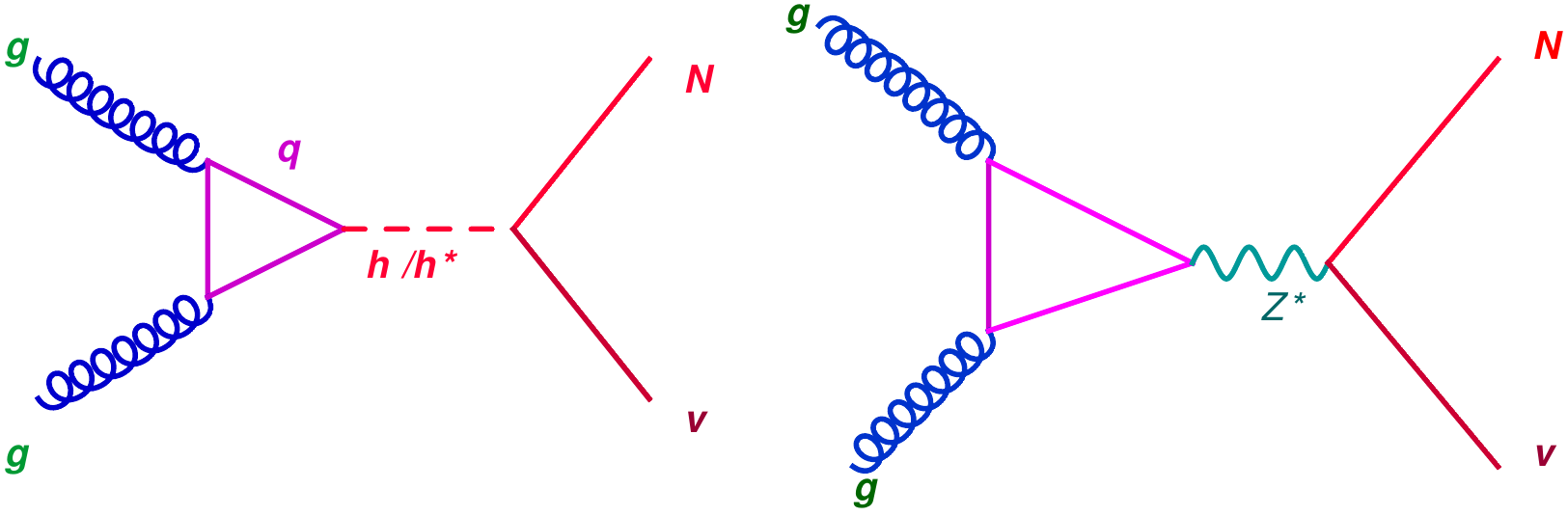}
\caption{ggF production of the heavy neutrinos from Higgs (h) and Z bosons. $h \to N \nu$ can be prompt when $m_h > m_N$ giving an enhancement to the $N\nu$ production from the ggF channel. For $m_h < m_N$, off-shell Higgs $(h^{\ast})$ produces $N\nu$ final state. Similarly off-shell $Z$ $(Z^{\ast})$ can also produce $N\nu$ final state when $m_{N} > m_Z$. For the Higgs mediated case, top quark loop will contribute dominantly whereas for the $Z^\ast$ case all contributions coming from the quarks will be counted according to its coupling with the quarks.}
\label{fig:6}
\end{center}
\end{figure*}
The production cross sections at the $13$ TeV LHC and future $100$ TeV pp collider are shown in Figs.~\ref{fig:13} and \ref{fig:100} respectively as a function of $m_N$.
\item[(3)] Gluon-fusion channel (ggF): \\
In the ggF top loop produces the SM Higgs and Higgs can decay into $N$ in association with $\nu$ \cite{Das:2017zjc, Das:2017rsu, Hessler:2014ssa}. The corresponding Feynman diagram is given in the left panel of Fig.~\ref{fig:6}.
There is another complementary production channel of $N$ in association with $\nu$ which comes from the $Z$ \cite{Das:2017zjc,Ruiz:2017yyf}. The corresponding Feynman diagram is given in the right panel of Fig.~\ref{fig:6}
The production process depends upon  $m_{N}$, one can produce them promptly from the Higgs decay when $m_{N} < m_{h}$ where $m_{h}$ is the Higgs mass and also from the off-shell $Z$ $(Z^\ast)$. 
 When $m_N < m_h$, Higgs can decay into RHNs and the partial decay width can be written as 
 \bea
 \Gamma_{h}^{\rm{new}} = \frac{Y_{D}^{2} m_{H}}{8 \pi} (1-\frac{m_{N}^{2}}{m_{h}^{2}})^{2}.
\eea
This has additive contribution to the total decay width of the SM Higgs boson keeping $Y_{D}$ as a free parameter. Constraints  on $Y_{D}$ and $|V_{\ell N}|^2$ have recently been studied in \cite{BhupalDev:2012zg, Das:2017zjc, Das:2017rsu}.
The cross section goes down at the 13 TeV LHC for $m_h < m_N$ due to the off-shell decay. However, at the future $100$ TeV pp collider the cross section again rises around $m_{N}=250$ GeV due to large contributions coming from the gluon PDFs. In this channel we are testing $m_N= 100$GeV $-$ $1$ TeV. The production cross sections are shown in Figs.~\ref{fig:13} and \ref{fig:100} respectively with respect to $m_N$. At the future $100$ TeV, the large gluon PDF starts contributing 
to the RHN production resulting a rise in the ggF curve compared to the 13 TeV result.
\item[(4)] Photon initiated processes:\\
The photon (denoted by ${\tt a}$ in {\tt MadGraph)} initiated processes also have important contributions in the RHNs production which have been studied in \cite{Williams:1934ad, vonWeizsacker:1934nji, Budnev:1974de, Dev:2013wba, Alva:2014gxa, Das:2015toa,  Deppisch:2015qwa}. In case of the the photon initiated process, the photon can be radiated from the proton and also from a parton. Those production channels will give additive contributions at the colliders once the RHNs decay into multi-lepton modes. The corresponding  Feynman diagrams for these processes are given in Figs.~\ref{fig:8}. The $N\ell j$ production process is possible as $\gamma \to WW$ vertex is present which will make the production cross section for $N\ell j$ higher in comparison to that of $N\nu j$ where $\gamma \to ZZ$ vertex is not present under the SM gauge group. $N\nu j$ will be produced from the $Z\to N \nu$ vertex under the NC interaction according to Eq.~\ref{NC}. For the photon initiated production processes we use\\
{\tt > generate ${\tt p a  \to N \ell j ~QED=3~QCD=0}$}\footnote{QED (Quantum Electrodynamic) processes switches the QED interactions with photon, $W$ and $Z$ bosons, QED$=3$ means the order of the interaction whereas QCD stands for the Quantum Chromodynamic processes with QCD$=0$ is the order of the interaction.}\\
with $p_T^{j} > 30$ GeV, $p_{T}^{\ell} > 10$ GeV, $|\eta^{\ell, j}| < 2.5$. To generate the $N\nu j$ final state with photon initiated process we use\\
{\tt > generate ${\tt p a  \to N \nu j ~QED=3~QCD=0}$}\\
with $p_T^{j} > 30$ GeV, $|\eta^{j}| < 2.5$. 
\begin{figure}
\begin{center}
\includegraphics[scale=0.6]{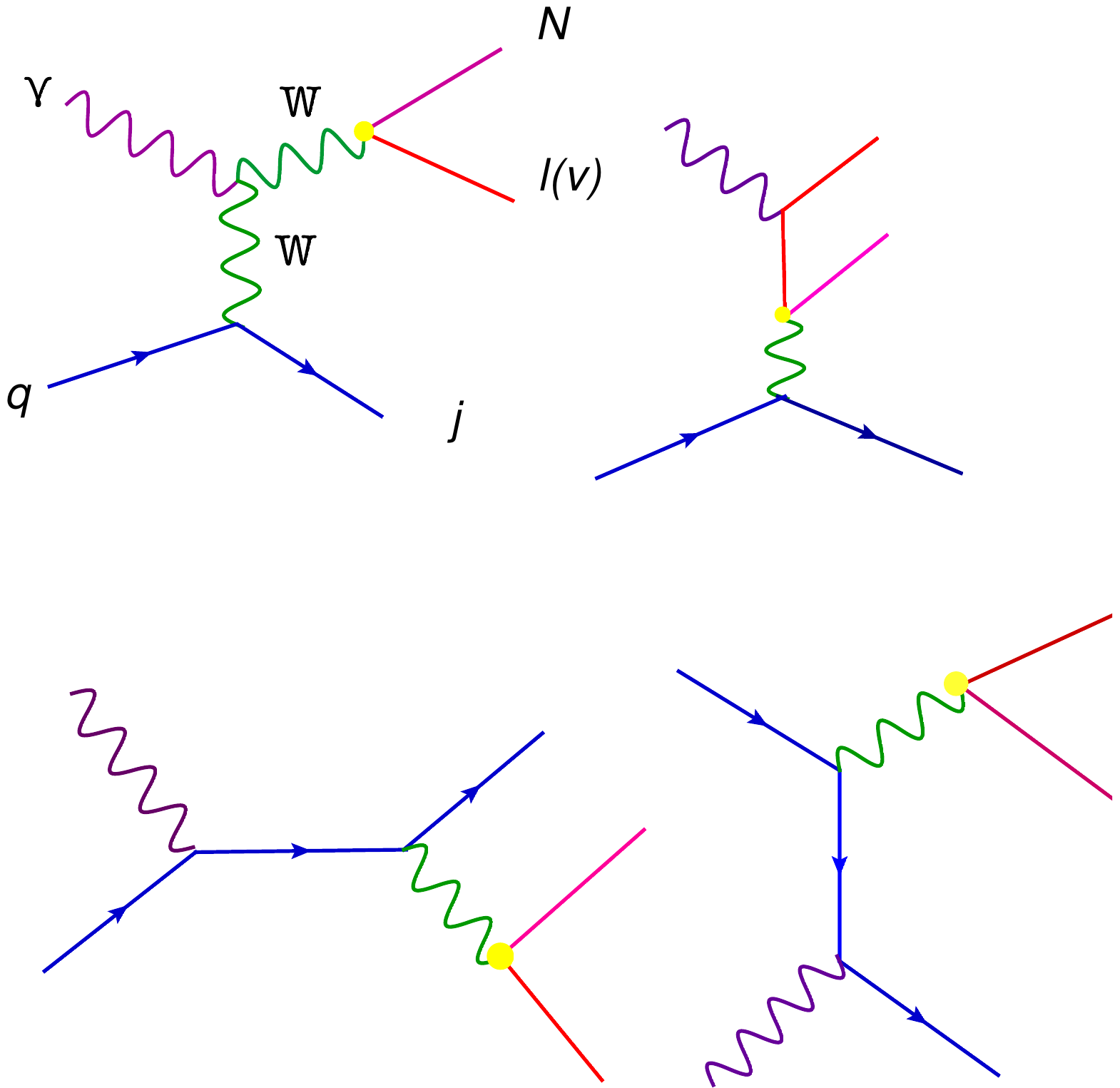}
\end{center}
\caption{The photon initiated $N\ell j$ final state. The $N\nu j$ final state can also be obtained, however, the contribution will be low as $\gamma \to ZZ$ vertex is absent due to the gauge invariance.}
\label{fig:8}
\end{figure}
\begin{figure}
\begin{center}
\includegraphics[scale=0.3]{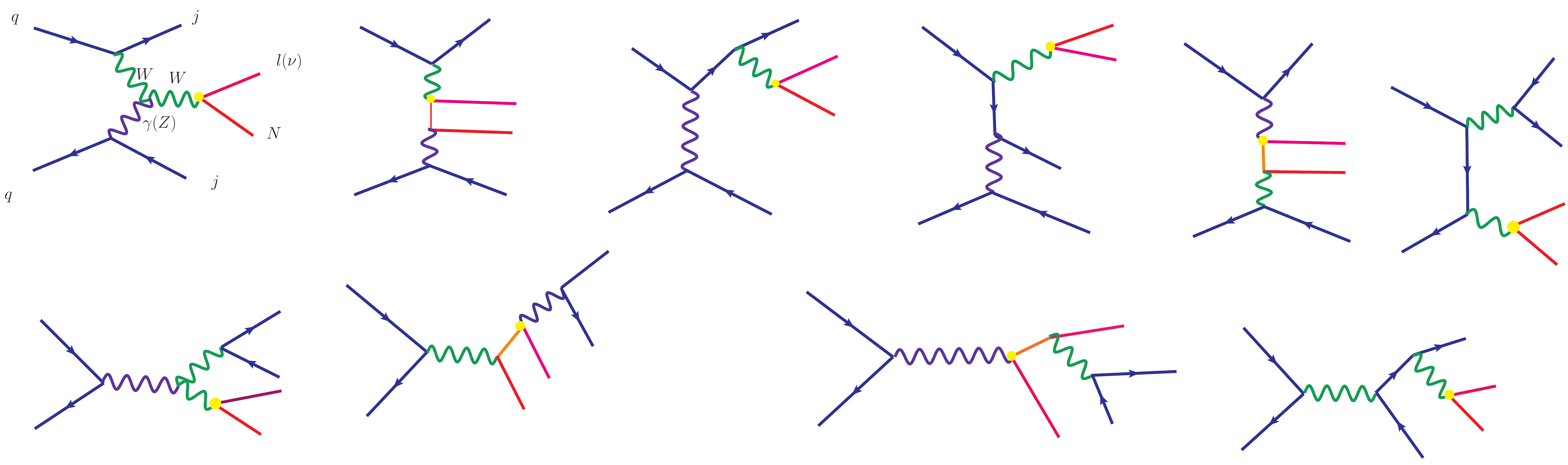}
\end{center}
\caption{VBF production of the $N\ell jj$ final state from the $W \to N\ell$ vertex. The $N\nu jj$ finals state can also be obtained using the $Z\to N\nu$ vertex. The two jets can be obtained at the forward- backward regions.}
\label{fig:9}
\end{figure}
\item[(5)] Vector Boson Fusion (VBF) channels:\\
The VBF processes are shown in Fig.~\ref{fig:9} where the $N\ell$ and $N\nu$ are produced with two widely separated jets and the corresponding final states will be $N\ell j j (N\nu j j)$ depending upon the CC (NC) vertex.
In the VBF case the $\gamma WW$ and $ZWW$ vertices will come into effect to produce the $N$ in association with $\ell (\nu) $ depending upon the CC (NC) between $N-W (N-Z)$. To generate the $N\ell jj(N\nu j)$ final state 
from the VBF channels we use\\
{\tt >generate ${\tt p p \to N \ell j j~QED=4~QCD=0}$}\\
with $p_T^{j} > 30$ GeV, $p_{T}^{\ell} > 10$ GeV, $|\eta^{\ell}| < 2.5$, $|\eta^{j_{1}}-\eta^{j_{2}}| > 4.2$,  $p_T^{j_{1},^{\rm{leading}}}=p_T^{j_{2},^{\rm{trailing}}} > 50$ GeV, $m_{jj}^{min}=250$ GeV.\\
{\tt >generate ${\tt p p \to N \nu j j~QED=4~QCD=0}$}\\
with $p_T^{j} > 30$ GeV, $p_{T}^{\ell} > 10$ GeV, $|\eta^{\ell}| < 2.5$, $|\eta^{j_{1}}-\eta^{j_{2}}| > 4.2$,  $p_T^{j_{1},^{\rm{leading}}}=p_T^{j_{2},^{\rm{trailing}}} > 50$ GeV, $m_{jj}^{min}= 250$ GeV.
The corresponding production cross sections for the 13 TeV LHC and future $100$ TeV pp collider are shown in Figs.~\ref{fig:13} and \ref{fig:100} respectively with respect to $m_N$.
\end{itemize}
\begin{figure*}
\begin{center}
\includegraphics[scale=0.43]{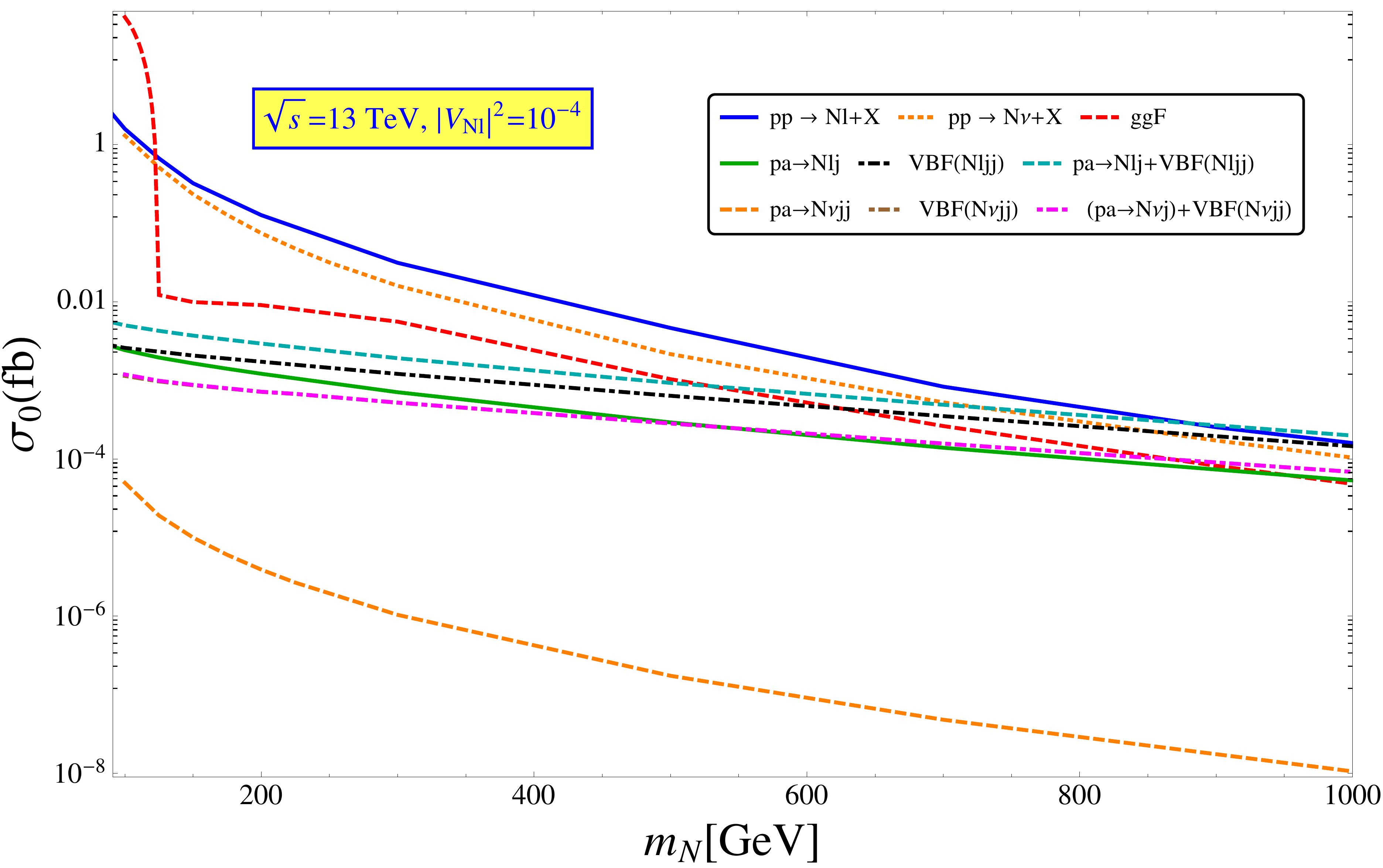}
\caption{Production cross  sections of the RHNs in association with leptons  from different modes at the 13 TeV LHC with $|V_{\ell N}|^2= 10^{-4}$. `a' stands for photon $(\gamma)$ as it is defined by MadGraph.}
\label{fig:13}
\end{center}
\end{figure*}
\begin{figure*}
\begin{center}
\includegraphics[scale=0.43]{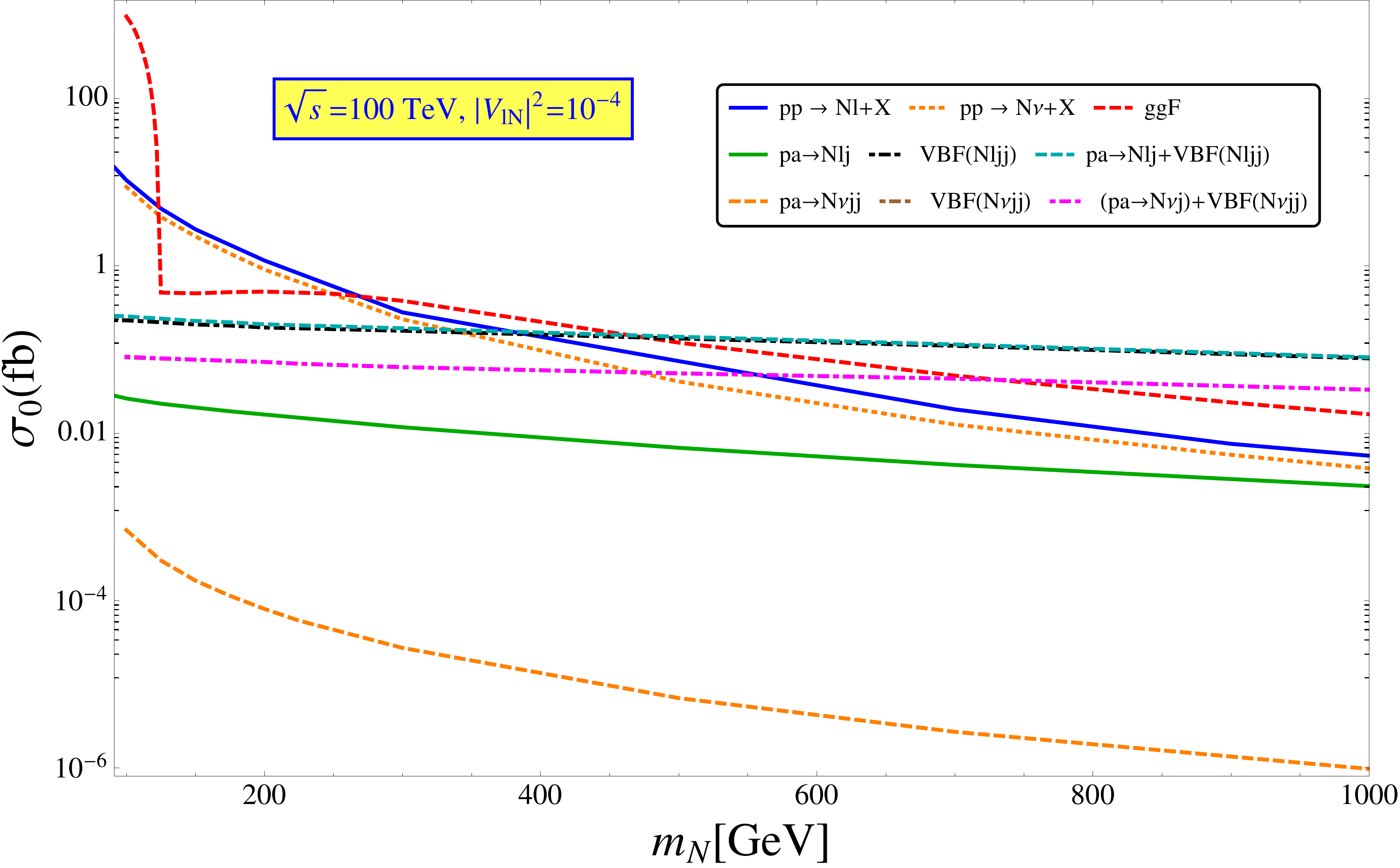}
\caption{Production cross  sections of the RHNs in association with leptons  from different modes at the future $100$ TeV pp collider with $|V_{\ell N}|^2= 10^{-4}$.`a' stands for photon $(\gamma)$ as it is defined by MadGraph.}
\label{fig:100}
\end{center}
\end{figure*}
\section{Decay of the RHNs at the LHC}
\label{decay}
\begin{figure}[h]
\includegraphics[scale=0.44]{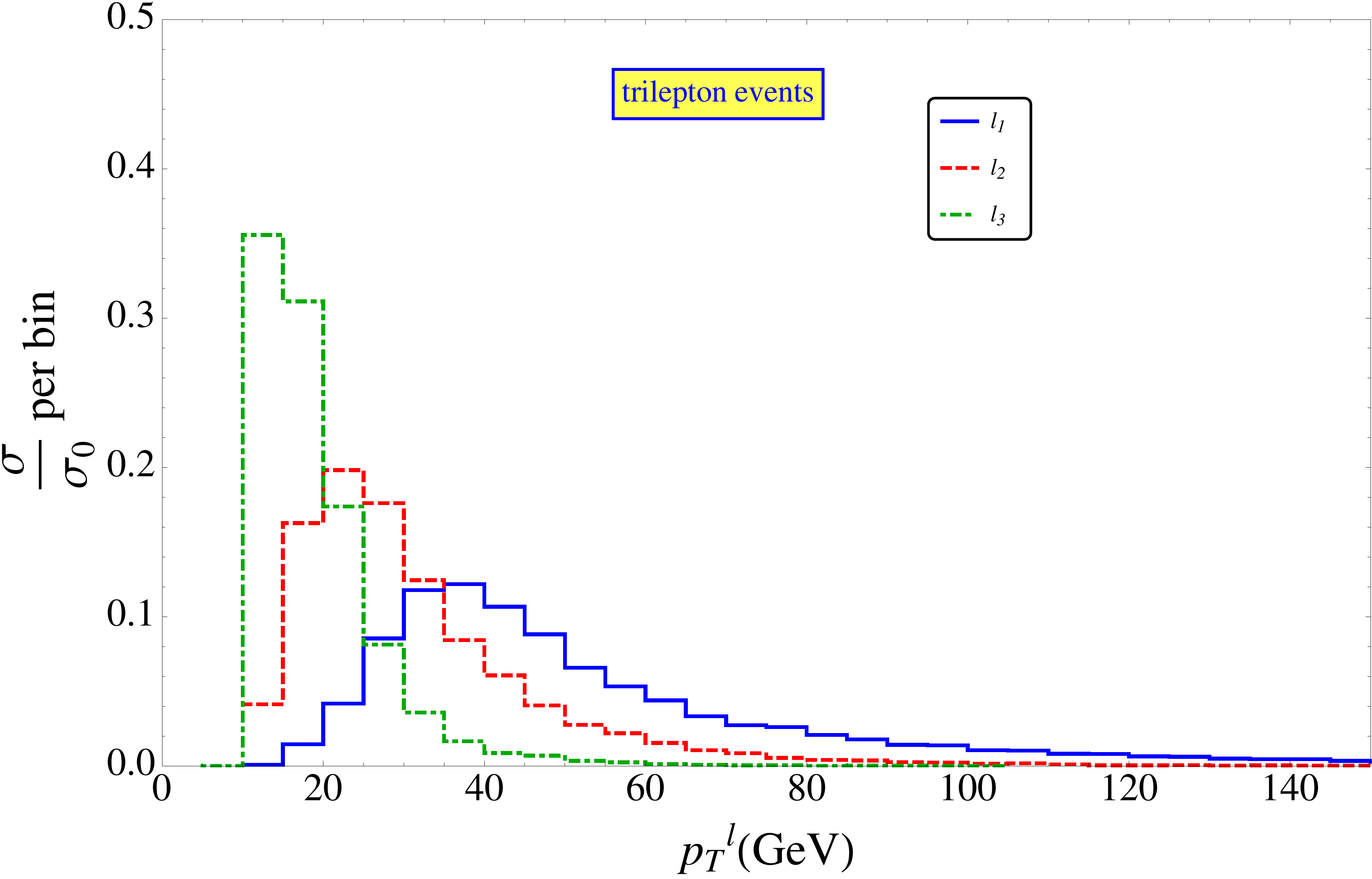}
\caption{Transverse momenta distributions for the three leptons from the $\tt{3\ell+MET}$ final state from $ pp\to N\ell X, N \to \ell W, W \to \ell \nu$ at the 13 TeV LHC.}
\label{Histo1}
\end{figure}

\begin{figure}[h]
\includegraphics[scale=0.44]{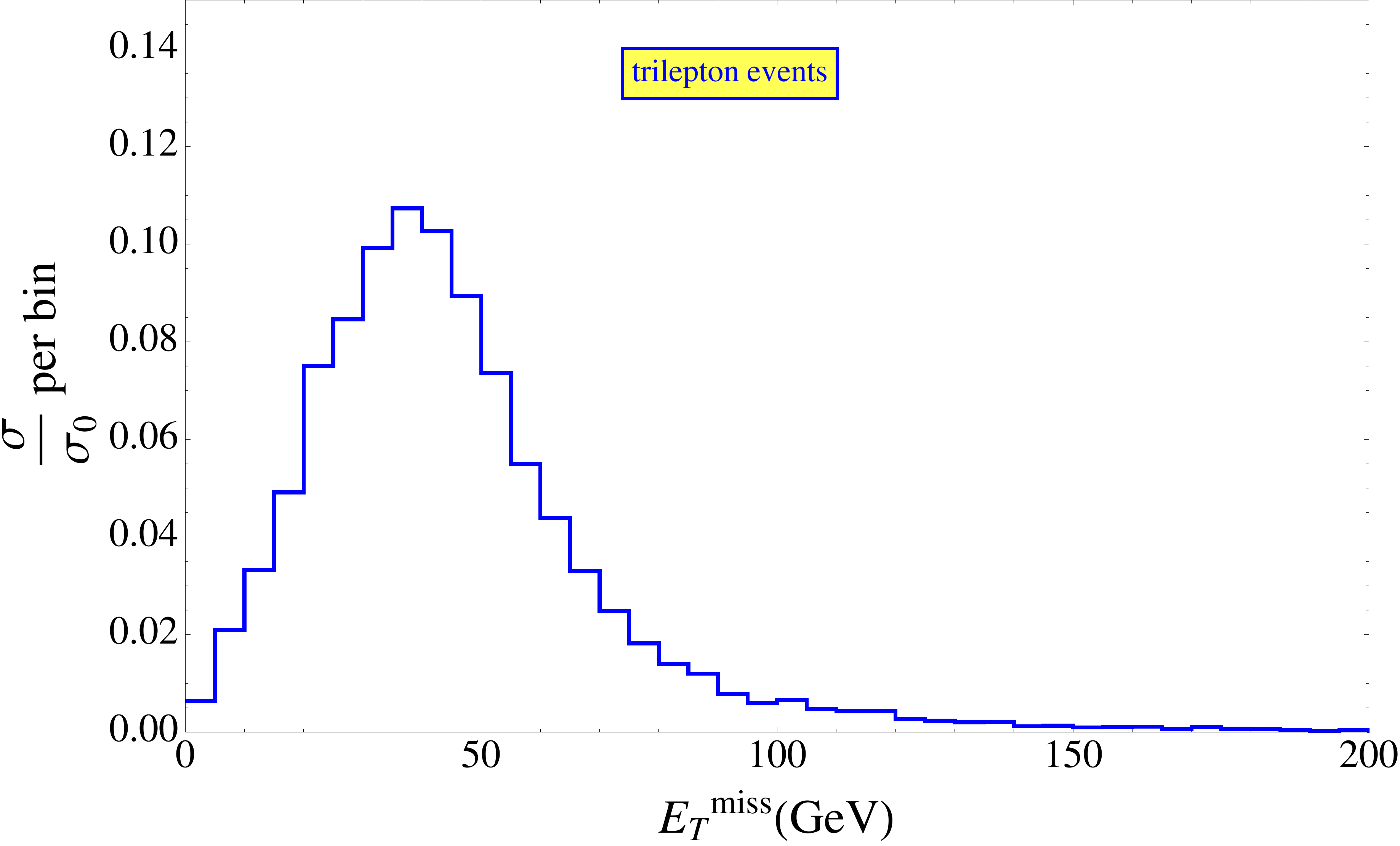}
\caption{MET distribution from the $\tt{3\ell+MET}$ final state from $ pp\to N\ell X, N \to \ell W, W \to \ell \nu$ at the 13 TeV LHC.}
\label{Histo12a}
\end{figure}

Implementing our model in the event generator {\tt MadGraph}, we produce the RHNs and allow it to decay in various multi-lepton modes. We use kinematic cuts in this adopting from the CMS analysis in Ref.\cite{CMS-trilep}. According to our choice we consider two models such as type-I Seesaw and  Inverse Seesaw. In type-I Seesaw the RHN is Majorana type and in inverse Seesaw the RHN is pseudo-Dirac type. From the Majorana RHN we get LNV same-sign di-lepton plus di-jet signal. Whereas from the pseudo-Dirac type RHNs we get LNC trilepton plus MET $(E_T^{miss}/ \slashed{E}_T)$ final state. Considering the leading decay mode of the RHNs $( N \to W \ell)$ and depending upon the choice of the final states, the $W$ will decay either leptonically or  hadronically. We consider the 13 TeV LHC to study such processes. We consider $m_{N}=100$ GeV in this section for different production processes. We use the data files after the hadronization using {\tt PYTHIA} \cite{Sjostrand:2006za} and detector simulation using {\tt DELPHES} \cite{deFavereau:2013fsa} bundled with MadGrpah \cite{Alwall:2011uj} and use the MLM matched results as prescribed in the previous section. In this analysis the leading particle is the particle having longest transverse momentum $(p_T)$ distribution where as the trailing particle is that which has shorter transverse momentum $(p_T)$ distribution. 
We have plotted the transverse momentum distributions of the leptons $(p_{T}^{\ell})$ in Fig.~\ref{Histo1} and the Missing Transverse Energy (MET/$E_{T}^{miss}/ \slashed{E}_T$) distribution in  the Fig.~\ref{Histo12a} from the trilepton plus MET $(3\ell+$MET$+X)$ signal. We notice that the leading lepton transverse momentum, $p_{T}^{\ell}$ for $\ell_1$ peaks around $40$ GeV where as the trailing leptons mostly stay between $10$ GeV$-40$ GeV. Looking at the distribution of the leading lepton and the peak of the distribution we suggest that the leading lepton is coming from the leptonic decay of the $W$. The MET distribution peaks around $40$ GeV which also backs the ides of having the leading lepton of the $W$. Therefore for the trilepton analysis at $m_N=100$ GeV using $E_{T}^{miss} < 50$ GeV will be reasonable choice.

\begin{figure}[h]
\includegraphics[scale=0.44]{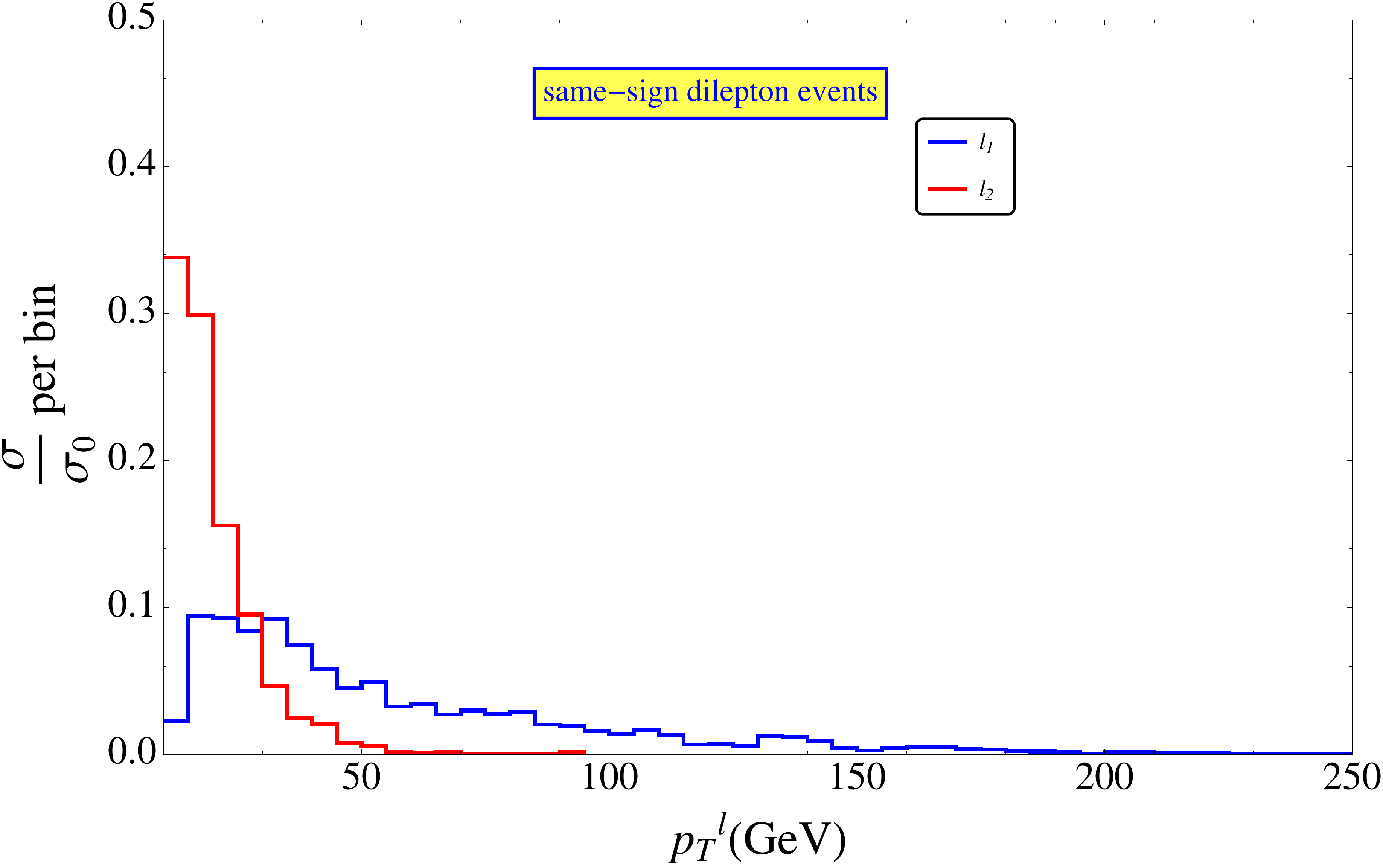}
\caption{Transverse momentum distributions of leptons $(p_{T}^{\ell})$ from the same-sign di-lepton plus di-jet final state events.}
\label{Histo4}
\end{figure}

\begin{figure}[h]
\includegraphics[scale=0.44]{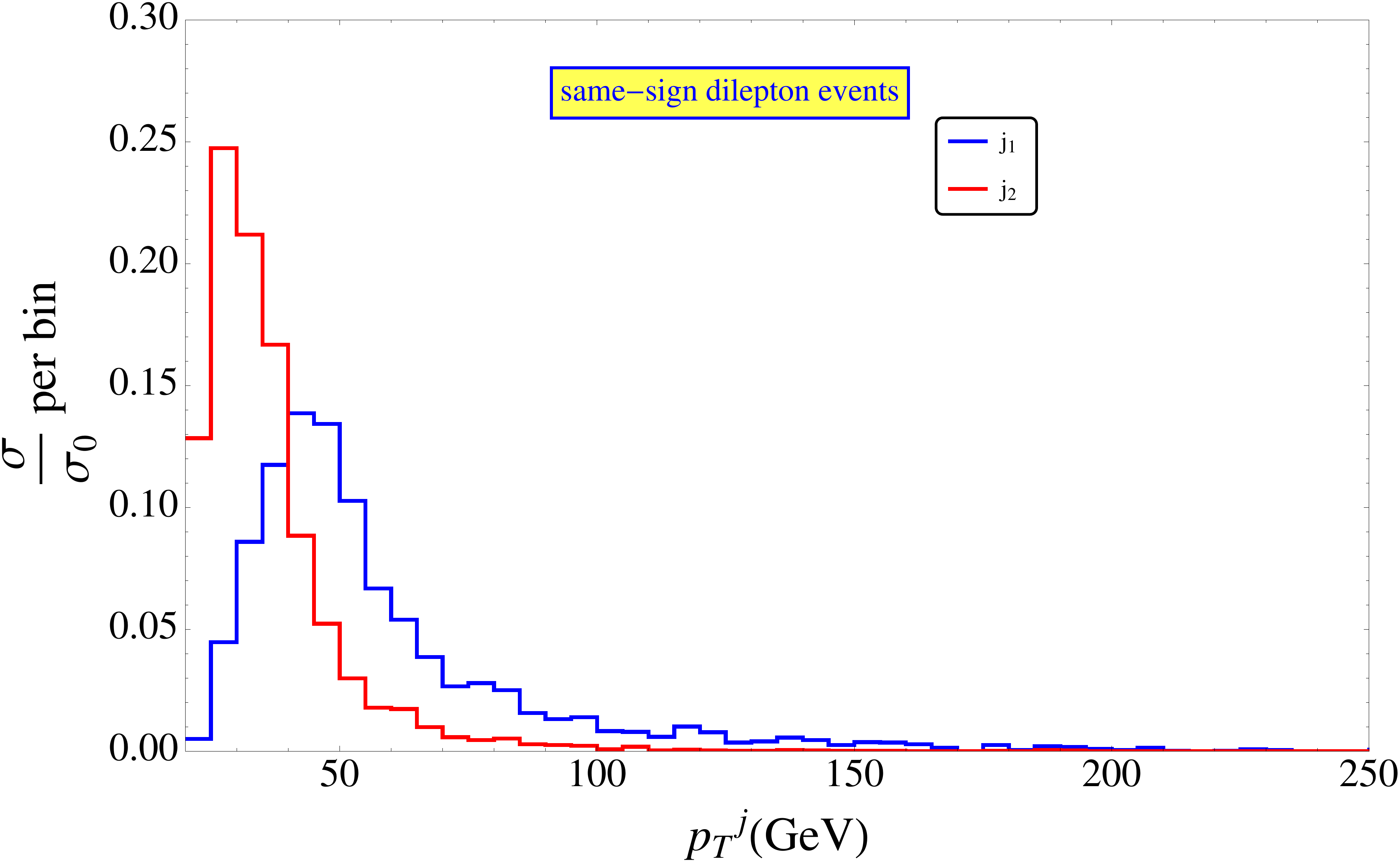}
\caption{Transverse momentum distributions of the jets $(p_{T}^{j})$ from the same-sign di-lepton plus di-jet final state. }
\label{Histo4a}
\end{figure}

\begin{figure}[h]
\includegraphics[scale=0.44]{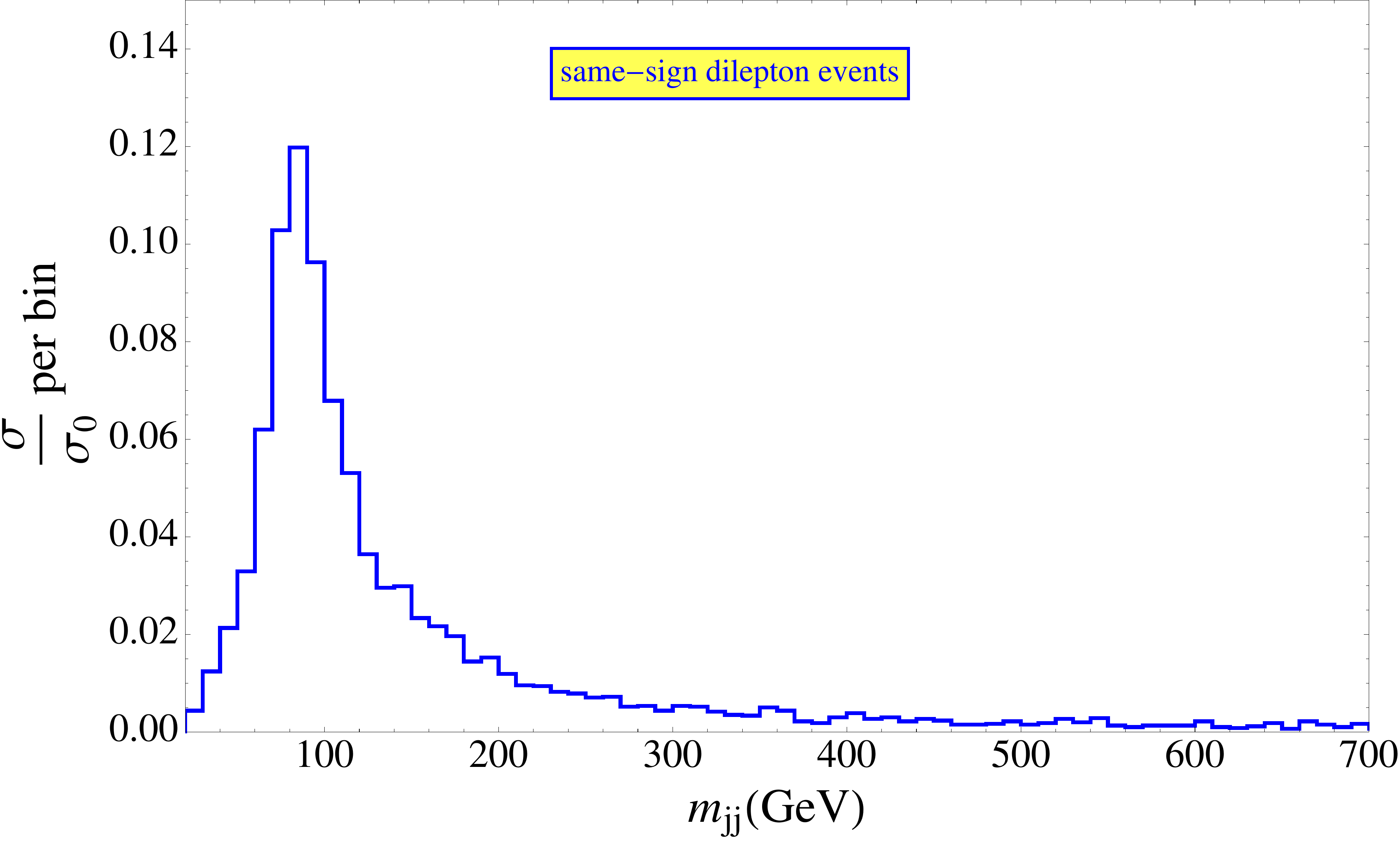}
\caption{The invariant mass distributions of two-jet $(m_{jj})$  system from the same-sign di-lepton plus di-jet final state showing a peak around the $W$ mass $(m_{W})$.}
\label{Histo4b}
\end{figure}

\begin{figure}[h]
\includegraphics[scale=0.44]{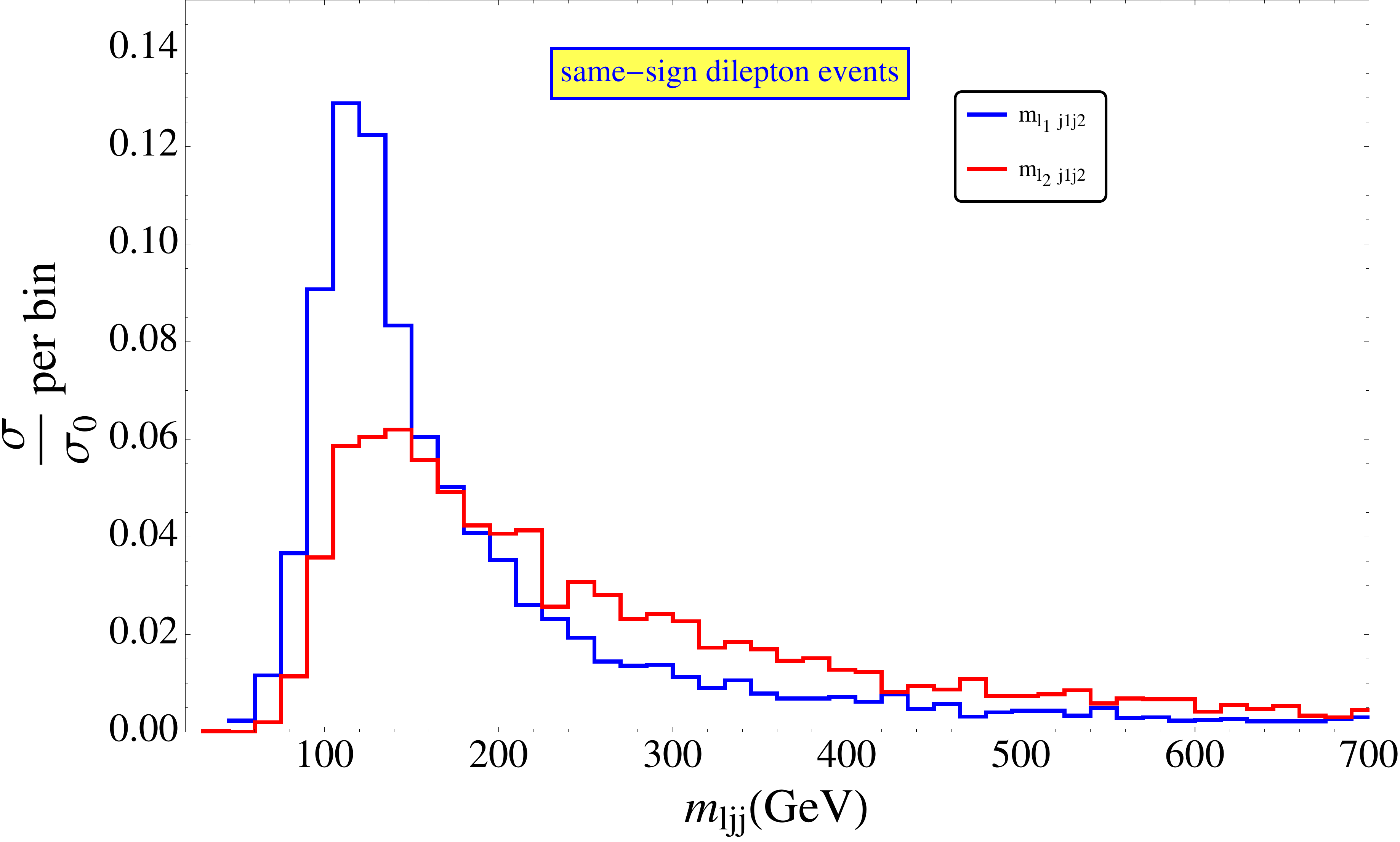}
\caption{The invariant mass distributions of one lepton plus two-jet $(m_{ljj})$ system showing a peak around $m_{N} \sim 100$ GeV (blue) from the same sign di-lepton plus di-jet signal.}
\label{Histo4c}
\end{figure}

We study the same sign di-lepton plus di-jet $(\ell^{\pm}\ell^{\pm} jj+X)$ where $X$ stands for the radiated jets. This signal is an important signature to test the Majorana nature of the RHNs produced from the type-I Seesaw model. 
The transverse momentum of the leptons $(p_{T}^{\ell})$ and the jets $(p_T^{j})$ are plotted in Figs.~\ref{Histo4} and \ref{Histo4a}, respectively. The invariant mass distribution of the two jets are plotted in 
Fig.~\ref{Histo4b}. The distribution shows a peak around $m_W$. The invariant mass distribution of the lepton plus two-jet system has been shown in Fig.~\ref{Histo4c}. The figure show the distributions including 
$m_{\ell_{1}jj}$ and $m_{\ell_{2}jj}$. Among them $m_{\ell_{1}jj}$ shows a peak around the assumed $m_{N}$ value around $100$ GeV. 

\begin{figure}[h]
\includegraphics[scale=0.54]{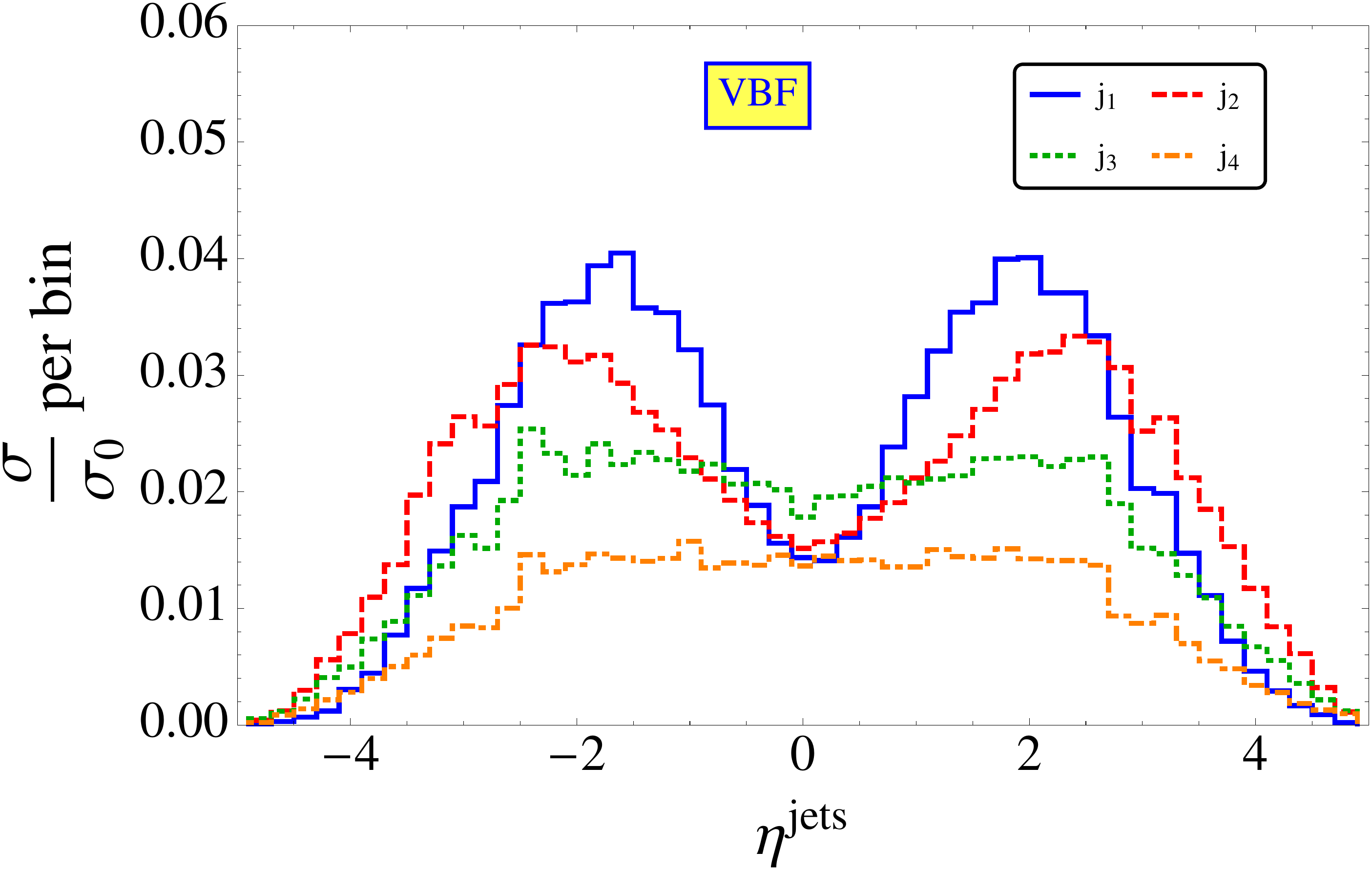}
\caption{Rapidity distributions of the associated jets in the trilepton mode from VBF process.}
\label{Histo2}
\end{figure}

\begin{figure}[h]
\includegraphics[scale=0.54]{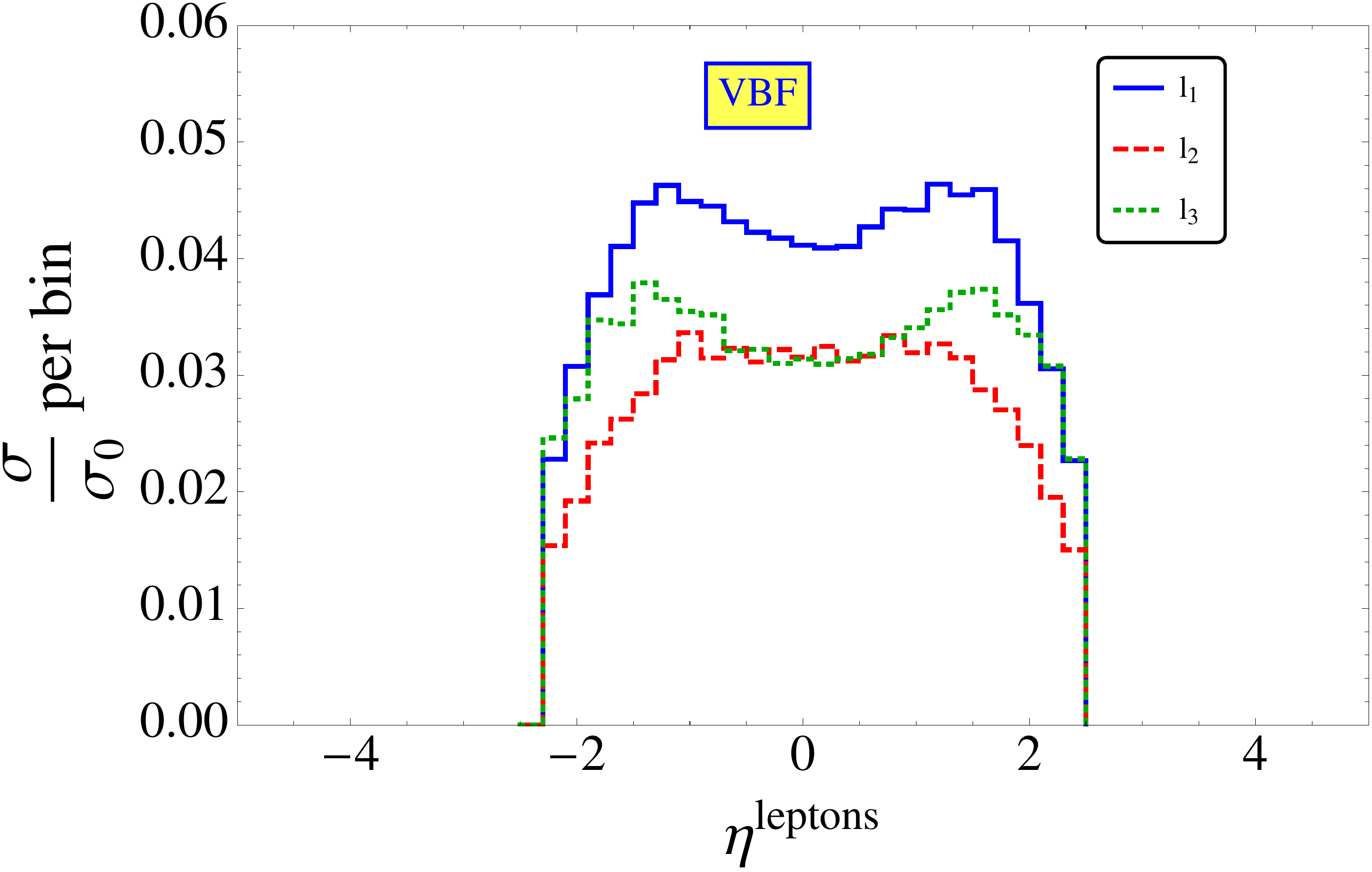}
\caption{Rapidity distributions of the the leptons in the trilepton mode from VBF process.}
\label{Histo2a}
\end{figure}

We study the VBF processes to produce the heavy neutrinos and hence study the trilepton plus MET final state. We have used the VBF prescriptions written in the previous section where the jets $j_{1}$ and $j_{2}$ are widely separated.
The other two jets are initial state radiations (ISR) which also are not populating the central region.  The rapidity distributions of the jets are shown in Fig.~\ref{Histo2}. This is a striking feature of the VBF process.
 The leading lepton coming out of the $W$ decay after the $W$ boson is produced from the $N\to W\ell$ decay. The leading leptons are also slightly away form the central region. We use a lepton pseudo-rapidity cut, $|\eta^{\ell}|<2.5$. The rapidity distributions of the leptons are shown in the Fig.~\ref{Histo2a}.


\begin{figure}[h]
\includegraphics[scale=0.45]{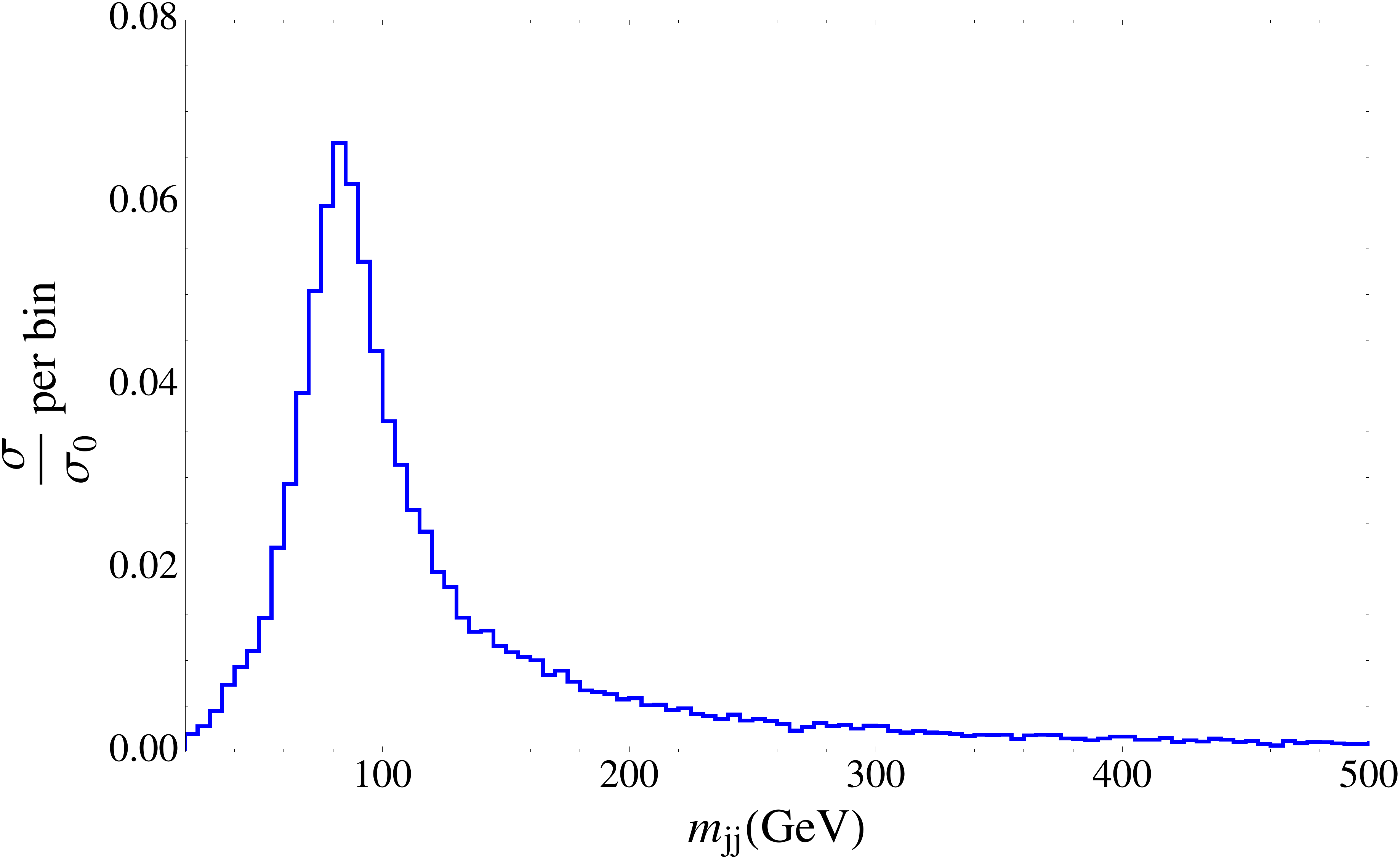}
\caption{Invariant mass distribution of the two jet $(m_{jj})$ from the $\ell+MET$$+2j+X$ final state from the $pp\to N\nu X$ process. 
The two jet system shows a peak around the $W$ mass $(m_{W})$.}
\label{Histo3a}
\end{figure}

\begin{figure}[h]
\includegraphics[scale=0.45]{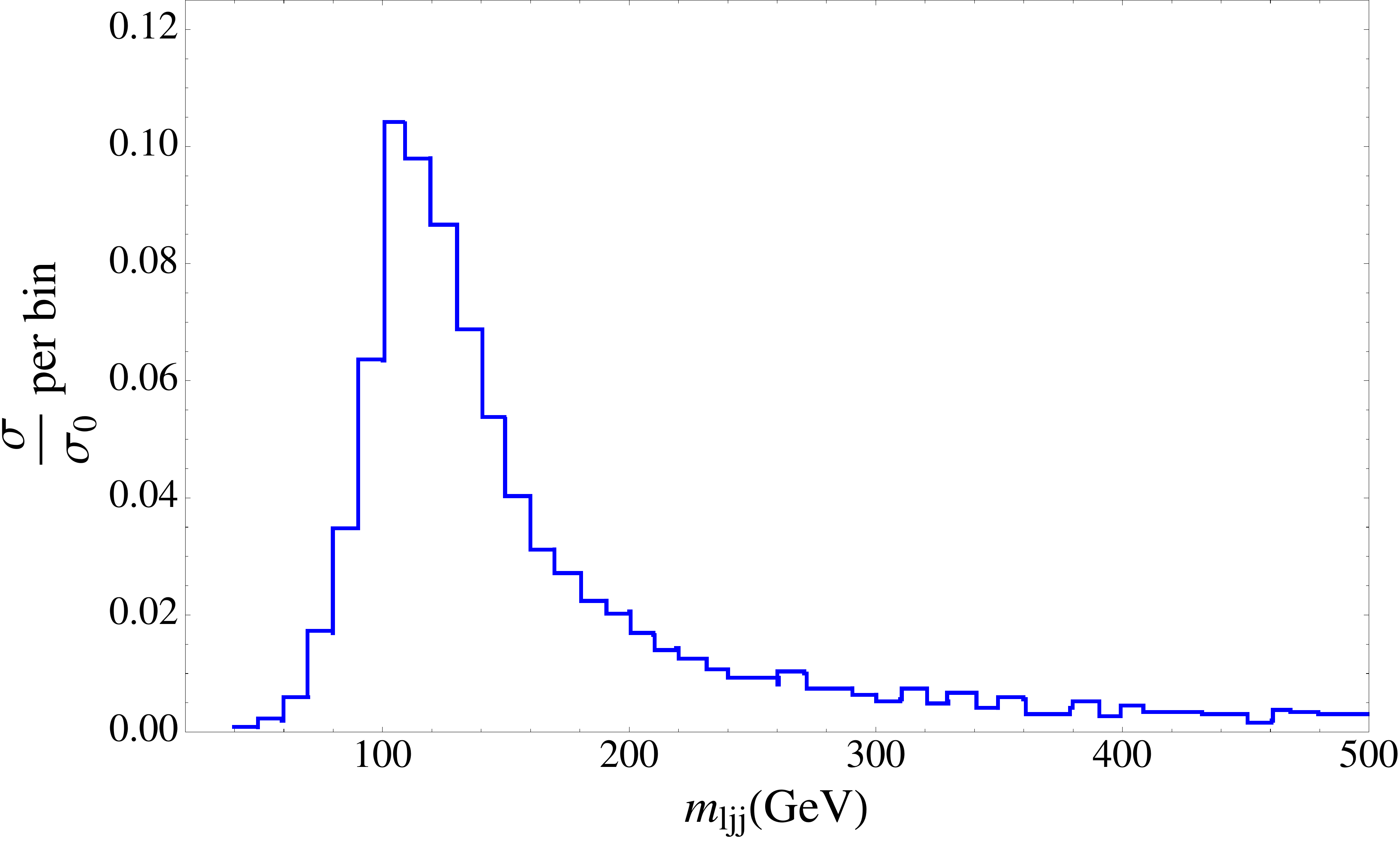}
\caption{Invariant mass distribution of the lepton plus two jet systems $(m_{\ell jj})$ from the $\ell+MET$$+2j+X$ final state from the $pp\to N\nu X$ process. 
The lepton plus two jet system shows a peak around $m_N\sim 100$ GeV.}
\label{Histo3b}
\end{figure}

Another important contribution is coming from the $N\nu j j+X$ process from the NC interaction between the RHNs and $Z$ bosons. Here $X$ are the radiated jets.
The final state consists of a single lepton, two jets and MET. The produced jets can be used to reconstruct the $W$ boson. The distribution shows a peak around the $W$ mass. Similarly from the 
RHN decay we get $N \to \ell j j$ and the invariant mass distribution of the $\ell jj$ system can show a peak around the RHN mass $(m_N)$. The invariant mass distribution of $m_{jj}$ is shown 
in Fig.~\ref{Histo3a} and that of $m_{\ell jj}$ are shown in Fig.~\ref{Histo3b}.


\begin{figure}[h]
\includegraphics[scale=0.43]{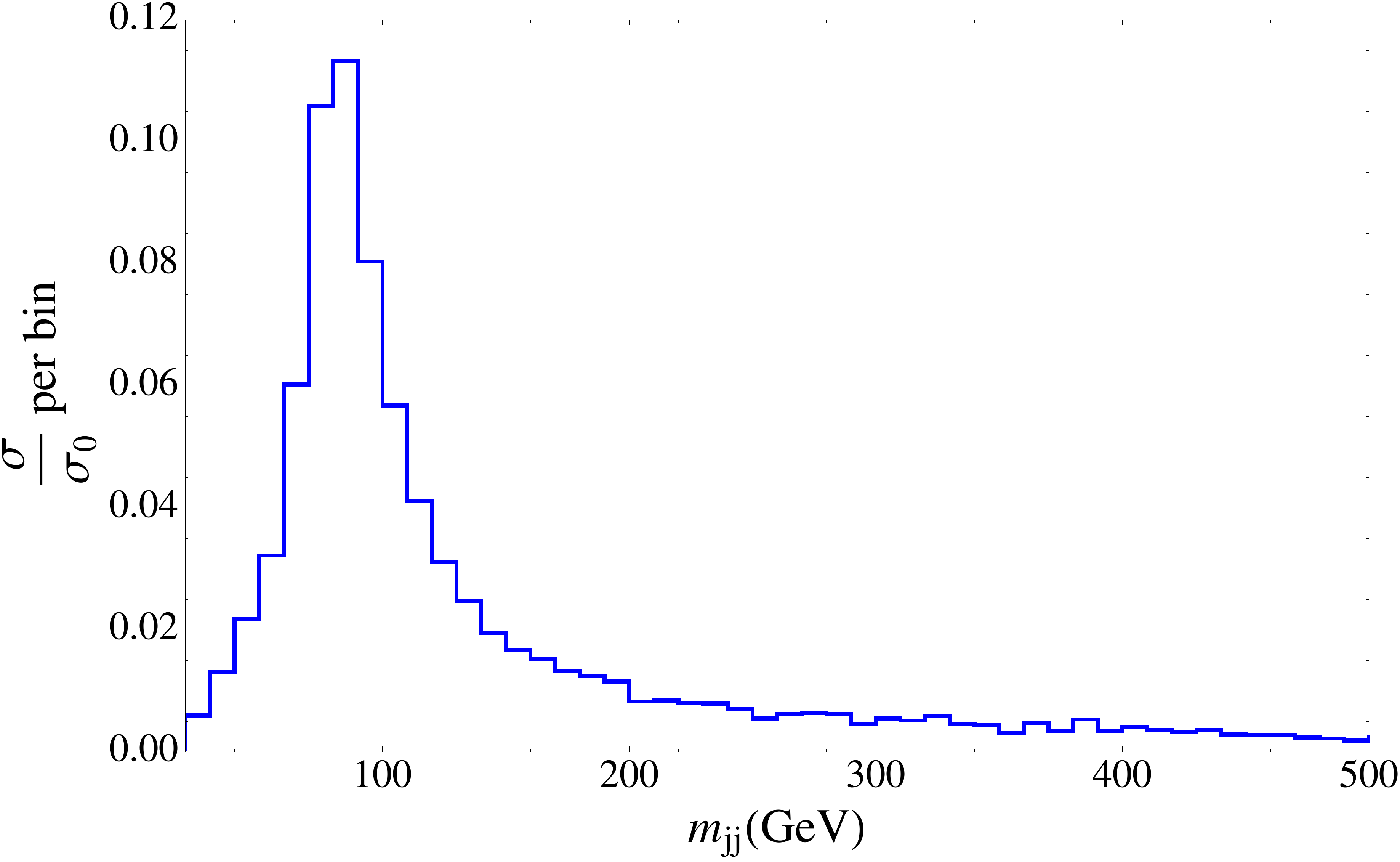}
\caption{Invariant mass distribution of the two jet $(m_{jj})$ from the $\ell+MET+2j$ final state coming from ggF  process.  The two jet system shows a peak around the $W$ mass $(m_{W})$.}
\label{Histo5}
\end{figure}

\begin{figure}[h]
\includegraphics[scale=0.43]{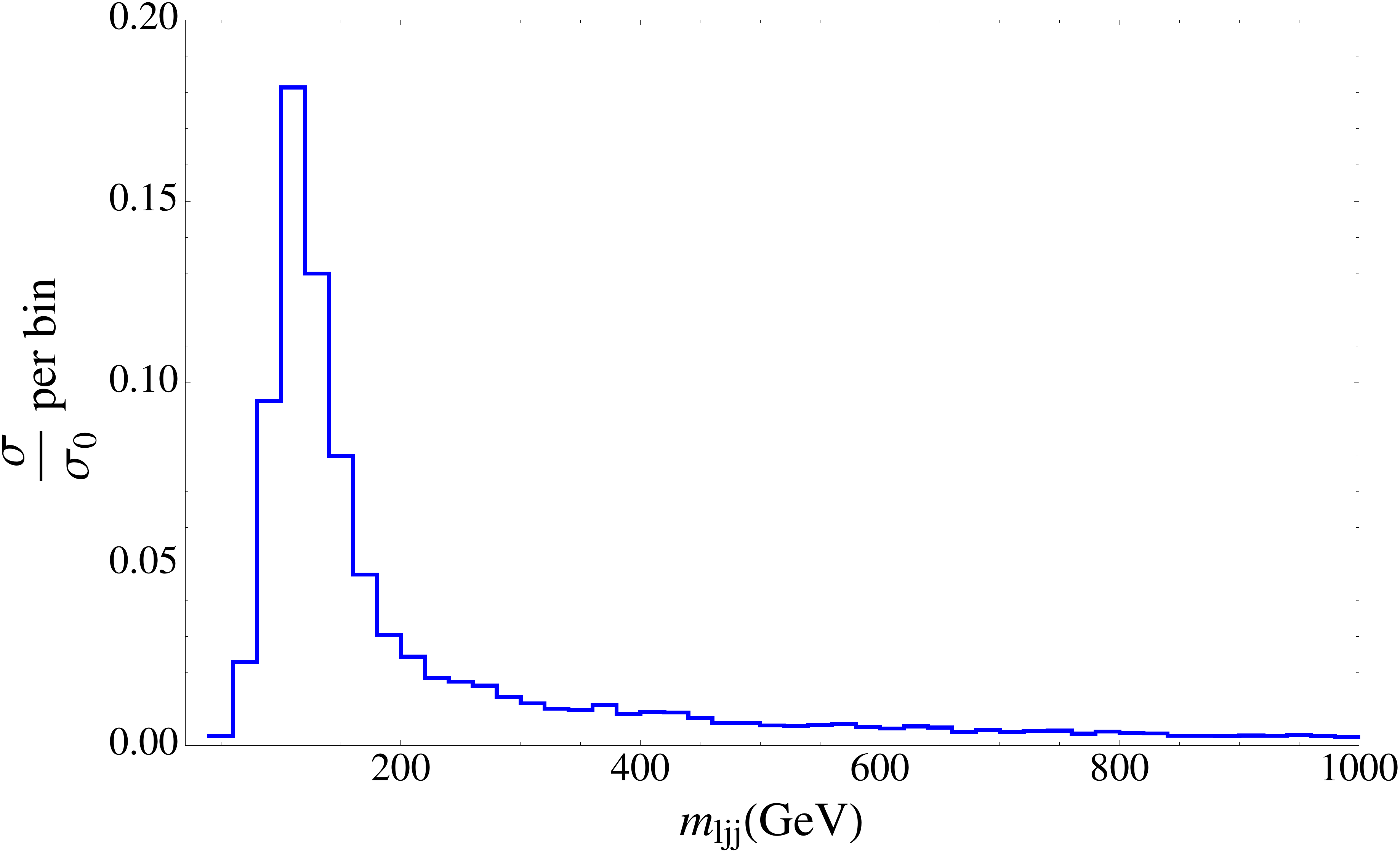}
\caption{Invariant mass distribution of the lepton plus two jet systems $(m_{\ell jj})$ from the $\ell+MET+2j$ final state coming from ggF  process. The lepton plus two jet system shows a peak around the RHN mass $(m_{N})$ at the 13 TeV LHC.}
\label{Histo5a}
\end{figure}

We produced the RHNs from the  ggF process where the top loop becomes dominant to produce the Higgs which can promptly decay into the RHN when $m_N < m_H$. In our case
we consider the $m_N=100$ GeV and the Higgs will decay promptly into $N\nu$ and the RHN will decay into $W$ followed by its hadronic decay. The invariant mass distribution
of the  two jet system $(m_{jj})$ is shown in Fig.~\ref{Histo5} where as that of the lepton and two jet system are shown in Fig.~\ref{Histo5a}. The $m_{jj}$ distribution
peaks around the $W$ mass where as the $m_{\ell jj}$ distribution peaks around the RHN mass. For detailed results regarding the Higgs decaying into RHNs can be found in \cite{ BhupalDev:2012zg, Das:2017zjc, Das:2017rsu}
with pervious and updated limits.

\section{Bounds on the mixing angle}
\label{bounds}
For $m_N < M_Z$, the RHN can be produced from the on-shell $Z$-decay through the NC interaction in association with missing energy, however, if $m_N > M_Z$ off-shell production will take place with the same final state. The heavy neutrino can decay according to the CC and NC interactions. Different production processes, different decay modes of the heavy neutrinos and various phenomenologies have been discussed in \cite{ Drewes:2013gca, Dube:2017jgo, Das:2017hmg, Das:2017gke,Bhardwaj:2018lma, Das:2017flq,  Das:2017pvt, Das:2016hof, Das:2017deo, Alva:2014gxa, Mitra:2016kov,Das:2017kkm, Drewes:2016jae, Tang:2017plx}. RHN production from various initial states and as well as  scale dependent production cross sections at the Leading Order (LO) and Next-to-Leading-Oder QCD (NLO QCD) of $N\nu$ have been studied at the 14 TeV LHC and future $100$ TeV pp collider \cite{Das:2016hof,Das:2017pvt}. The L3 collaboration \cite{Adriani:1992pq} has performed a search on such heavy neutrinos directly from the LEP data and found a limit on $\mathcal{B}(Z\to \nu N)< 3\times 10^{-5}$ at the $95\%$ CL for the mass range up to 93 GeV. The exclusion limits from L3 are given in Fig.~\ref{mix} where the red dot-dashed line stands for the limits obtained from electron (L3-$e$) and the red dashed line stands for the exclusion limits coming from $\mu$ (L3-$\mu$). The corresponding exclusion limits on $|V_{(\ell=e) N}|$ at the $95\%$ CL  \cite{Acciarri:1999qj,Achard:2001qv} have been drawn from the LEP2 data which have been denoted by the dark magenta line. In this analysis they searched for $80~\rm{GeV} \leq M_{\it N} \leq 205~\rm{GeV}$ with a center of mass energy between 130 GeV to 208 GeV \cite{Achard:2001qv}.

The DELPHI collaboration \cite{Abreu:1996pa} had also performed  the same search from the LEP-I data which set an upper limit for the branching ratio $\mathcal{B} (Z\to N \nu)$ about $1.3\times 10^{-6}$ at $95\%$ CL for $3.5$ GeV $\leq m_{N} \leq 50$ GeV. Outside this range the limit starts to become weak with the increase in $m_N$.  Here it has been considered that $N\to W\ell$ and $N\to Z \nu$ decays after the production of the heavy neutrino was produced. The exclusion limits for $\ell=e, \mu, \tau$ are depicted by the blue dotted, dashed and dot-dashed lines in Fig.~\ref{mix}. 

The search of the sterile neutrinos can be made at high energy lepton colliders with a very high luminosity such as Future Circular Collider (FCC) for the Seesaw model. A design of such collider has been launched recently where nearly 100 km tunnel will be used to study high luminosity $e^+e^-$ collision (FCC-$ee$) with a center-of-mass energy around 90 GeV to 350 GeV \cite{Blondel:2014bra}. According to this report, a sensitivity down to $|V_{\ell N}|^{2}\sim10^{-11}$ could be achieved from a range of the heavy neutrino mass, $10$ GeV $\leq m_{N} \leq 80$ GeV. The darker cyan-solid line in Fig.~\ref{mix} shows the prospective search reaches by the FCC-$ee$. A sensitivity down to a mixing of $|V_{\ell N}|^2 \sim 10^{-12}$ can be obtained in FCC-$ee$ \cite{Blondel:2014bra}, covering a large phase space for $m_{N}$ from $10$ GeV to $80$ GeV.

The heavy neutrinos can participate in many electroweak (EW) precision tests due to the active-sterile couplings. For comparison, we also show the $95\%$ CL indirect upper limit on the mixing angle, $|V_{\ell N}| < 0.030, 0.041~\rm{and}~0.065$ for $ \ell=\mu, e, \tau$ respectively derived from a global fit to the electroweak precision data (EWPD), which is independent of $m_N$ for $ m_N > M_Z$, as shown by the horizontal pink dash, solid and dolled lines respectively in Fig.~\ref{mix}. The bounds from the EWPD have been studied in Refs.\cite{deBlas:2013gla, delAguila:2008pw, Akhmedov:2013hec}. For the mass range, $m_N < M_Z$, it is shown in \cite{Deppisch:2015qwa} that the exclusion limit on the mixing angle remains almost unaltered, however, it varies drastically at the vicinity of $m_N =1~\rm{GeV}.$ For the flavor universal case the bound on the mixing angle is given as $|V_{\ell N}|^{2}=0.025$ from \cite{deBlas:2013gla} which has been depicted in Fig.~\ref{mix} with a pink dot-dashed line.
 
 The CMS limits from the $8$ TeV LHC are represented by the Plink lines. The dashed Pink line represents the $\mu$ and solid pink line represents the $e$. For both of the flavors CMS has tested the mass range, $40$ GeV $\leq m_N \leq 200$ GeV. For electron it has probed down to $10^{-4}$ and for muon it has probed down to $10^{-5}$ for $|V_{e N}|^2$ and $|V_{\mu N}|^2$ respectively. The ATLAS bounds from the $8$ TeV  LHC are weaker than the CMS bounds. The ATLAS \cite{Aad:2015xaa} bounds are represented by the brown dashed lines for $\mu$ $(|V_{\mu N}|^2)$ and brown dotted line  represents the bounds from $e$ $(|V_{eN}|^2)$ in Fig.~\ref{mix}.
 
The relevant existing upper limits at the  $95\%$ CL are also shown to compare with the experimental bounds using the LHC Higgs boson data in \cite{BhupalDev:2012zg,Das:2014jxa} using \cite{Chatrchyan:2012ty, CMS:2012xwa,Aad:2012uub,Chatrchyan:2012ft,Aad:2012ora}. The darker green dot-dashed line named Higgs boson shows the relevant bounds on the mixing angle. In this analysis we will compare our results taking this line as one of the references. We have noticed that the $|V_{\ell N}|^2$ can be as low as $4.86\times10^{-4}$ while $m_N=60$ GeV and the bound becomes stronger at $m_N=100$ GeV as $3.73\times 10^{-4}$. When $m_N > 100$ GeV, the bounds on $|V_{\ell N}|^2$ become weaker.

\subsection{Same-sign di-lepton plus di-jet signal}

For simplicity we consider the case that only one generation of the heavy neutrino
  is light and accessible to the LHC which couples to only the second generation of the lepton flavor. 
To generate the events in the {\tt MadGraph} we use the {\tt CTEQ6L1 PDF} \cite{CTEQ}
with {\tt xqcut}$=p_{T}^{j}=20$ GeV and {\tt QCUT}$=25$ GeV.
We calculate the cross sections for the {\tt combined} processes $N\ell X, N \to \ell j j$
as functions of $m_{N}$ from various initial states 
as described in Fig.~\ref{fig:13} and \ref{fig:100}.
Comparing our generated events with the recent ATLAS results \cite{Aad:2015xaa} at the $8$ TeV LHC with the luminosity $20.3$ fb$^{-1}$,
  we obtain an upper limit on the mixing angles between the Majorana type 
  heavy neutrino and the SM leptons as a function of $m_{N}$.
In the ATLAS analysis the upper bound of the production cross section ($\sigma^{\rm{ATLAS/ CMS}}$) is obtained for the final state with
  the same-sign di-muon plus di-jet as a function of $m_{N}$. For $\sigma^{\rm{ATLAS}}$ results we use the cross sections given in \cite{Aad:2015xaa}
  and for $\sigma^{\rm{CMS}}$ we use \cite{CMS8}. Implementing our model in MadGraph we generate the signal event and compare the experimental 
  cross sections from ATLAS and CMS using 
\bea
|V_{\mu N}|^{2} \lesssim \frac{\sigma^{\rm{ATLAS/ CMS}}}{(pp\to N\ell X, N \to \ell j j)+ \rm{All~same~sign~di-lepton~modes} }.
\label{mix2j}
\eea

\begin{figure}[h]
\includegraphics[scale=0.45]{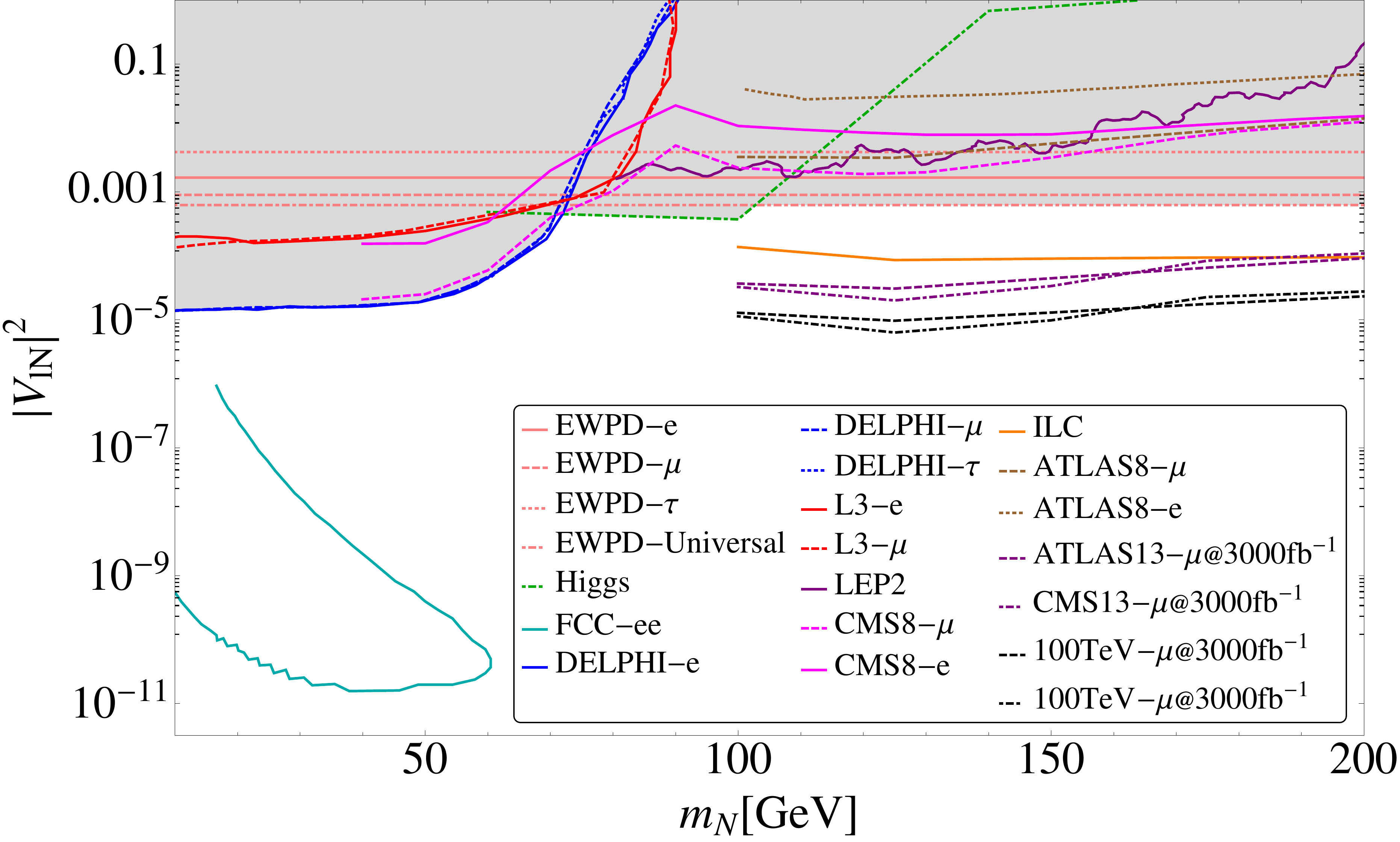}
\caption{Bounds on the square of the mixing angle from the same sign di-lepton plus di-jet final state as a function of the heavy neutrino mass $(m_N)$. The shaded region is ruled out by DELPHI, EWPD and Higgs data. Prospective bounds from the Higgs
data had been calculated in \cite{Das:2017zjc} for the LHC at the high luminosity $(3000$ fb$^{-1})$ and future $100$ TeV pp collider. }
\label{mix}
\end{figure}

Our upper bounds for the $13$ TeV LHC with $3000$ fb$^{-1}$ luminosity on the mixing angles are shown in Fig.~\ref{mix} along with the exclusion limits obtained from
  the different experimental results shaded in gray as described in the previous subsection.
We can see that a significant improvement on the bounds can be made adding the jets and various initial states as described in this paper.
Applying the ATLAS bound at $8$ TeV with $20.3$ fb$^{-1}$ luminosity \cite{Aad:2015xaa}, 
  we put a prospective upper bound on the mixing angles at the $13$ TeV LHC with $3000$ fb$^{-1}$ luminosity.
  Recently the CMS has performed the same-sign di-lepton plus di-jet search \cite{CMS8}.
Using this result and adopting the same procedure for the ATLAS result we calculate 
  the upper bound on the mixing angles at the $13$ TeV LHC with $3000$ fb$^{-1}$ luminosity using Eq.~\ref{mix2j}.
The results are shown in Fig.~\ref{mix}. 

A significant improvement for the upper bound on the mixing angle is obtained by combining all the processes with jets and various initial states as described in this paper.
We can see that at $91.2$ GeV$ \leq m_N \leq 200$ GeV,  the bounds obtained using the ATLAS $(\rm{ATLAS}13-\mu @3000fb^{-1})$  analysis is better than the recent exclusions limits shaded in gray.
Using the $8$ TeV CMS result \cite{CMS8} we obtain prospective bound at the 13 TeV LHC for the luminosity $3000$ fb$^{-1}$$(\rm{CMS}13-\mu @3000$ fb$^{-1})$ in Fig.~\ref{mix}.
The results are also shown in Fig.~\ref{mix}. In this analysis we have only chosen the $\mu$ flavor. The Purple lines represent the $13$ TeV LHC and black lines represent the future $100$ TeV pp collider. 
The corresponding improved bounds at the future $100$ TeV pp collider with $3000$ fb$^{-1}$ luminosity are shown in Fig.~\ref{mix} with the black dashed and 
dot-dashed lines corresponding to the ATLAS  and CMS data marked as $(100$TeV$-\mu @3000$fb$^{-1})$, respectively in Fig.~\ref{mix}. The prospective bounds calculated from Eq.~\ref{mix2j} at the $13$ TeV LHC are represented by the purple and those for the future $100$ TeV are represented by Black lines. 

We find that at the HL-LHC, using the same-sign di-muon and di-jet final state LHC can probe the squared of the mixing angle down to $1.95\times10^{-5}$ for the $\mu$ flavor where as the future $100$ TeV pp collider can probe the square of the mixing angle down to $5.83\times10^{-6}$ for the  $\mu$ flavor. 

\subsection{Trilepton signal}

In this analysis we consider two cases. 
One is the Flavor Diagonal case (FD), where
  we introduce three generations of the degenerate heavy neutrinos 
  and each generation couples with the single, corresponding lepton flavor.
The other one is the Single Flavor case (SF) where only one heavy neutrino
is light and accessible to the LHC which couples to 
only the first or second generation of the lepton flavor using the {\tt CTEQ6L PDF} \cite{CTEQ}.
In this analysis we consider $N\ell X$ final state followed by $N\to \ell \ell \nu_{\ell}$.

We generate the {\tt combined} parton level events  using {\tt MadGraph} and 
  then gradually hadronize and perform detector simulations with {\tt xqcut}$=p_{T}^{j}=30$ GeV and 
  {\tt QCUT}$=36$ GeV for the hadronization. 
In our analysis we use the matched cross section after the detector level analysis. 
After the signal events are generated we adopt the following basic criteria, 
  used in the CMS trilepton analysis \cite{CMS-trilep},
\begin{itemize}
\item [(i)] the transverse momentum of each lepton: $p^\ell_T > 10$ GeV;
\item [(ii)] the transverse momentum of at least one lepton: $p^{\ell,{\rm leading}}_{T} > 20$ GeV;
\item [(iii)]  the jet transverse momentum: $p_T^j > 30$ GeV;
\item [(iv)] the pseudo-rapidity of leptons: $|\eta^\ell| < 2.4$ and of jets: $|\eta^j| < 2.5$;
\item [(v)] the lepton-lepton separation: $\Delta R_{\ell \ell} > 0.1$ and the lepton-jet separation: $\Delta R_{\ell j} > 0.3$;
\item [(vi)] the invariant mass of each OSSF (opposite-sign same flavor) lepton pair: $m_{\ell^+ \ell^-}< 75$  GeV or $> 105$ GeV to avoid the on-$Z$ region which was excluded from the CMS search. Events with $m_{\ell^+ \ell^-}< 12$ GeV are rejected to eliminate background from low-mass Drell-Yan processes and hadronic decays;
\item [(vii)]  the scalar sum of the jet transverse momenta: $H_{T} < 200$ GeV;
\item [(viii)] the missing transverse energy: $\slashed{E}_{T} < 50$ GeV.
\end{itemize}
The additional trilepton contributions come from $N \rightarrow Z \nu, h \nu$, followed by the $Z$, $h$ decays into a pair of OSSF leptons. 
However, the $Z$ contribution is rejected after the implementation of the invariant mass cut for the OSSF leptons to suppress the SM background and 
  the $h$ contribution is suppressed due to very small Yukawa couplings of electrons and muons. 
Using different values of $\slashed{E}_{T}$ and $H_{T}$, the CMS analysis provides different number of observed events and 
  the corresponding SM background expectations. 
For our analysis the set of cuts listed above are the most efficient ones as implemented by the CMS analysis \cite{CMS-trilep}.
To derive the limits on $|V_{\ell N}|^{2}$, we calculate the signal cross section normalized by the 
  square of the mixing angle as a function of the heavy neutrino mass $m_{N}$
  for both SF and FD cases, by imposing the CMS selection criteria listed above. 
The corresponding number of signal events passing all the cuts 
  is then compared with the observed number of events at the 19.5 fb$^{-1}$ luminosity \cite{CMS-trilep}.
For the selection criteria listed above, the CMS experiment observed:
\begin{itemize}
\item[(a)] 510 events with the SM background expectation of 560$\pm$87 events for $m_{\ell^{+}\ell^{-}} <$ 75 GeV which is called below the $Z-$pole.
\item[(b)] 178 events with the SM background expectation of 200$\pm$35 events for $m_{\ell^{+}\ell^{-}} >$ 105 GeV which is called above the $Z-$ pole. 
\end{itemize}
In case (a) we have an upper limit of 37 signal events, while in 
  case (b) leads to an upper limit of 13 signal events. 
Using the limits obtained in case (b), we can set an upper bound on $|V_{\ell N}|^{2}$ for a given value of $m_{N}$. 
 In this analysis we use above the $Z$ pole situation with $3000$ fb$^{-1}$ luminosity for the 13 TeV LHC and future $100$ TeV pp collider.
 (Exactly the same procedure can be adopted for the situation with $m_{\ell^{+}\ell^{-}} <$ $75$ GeV.)

\begin{figure}[h]
\includegraphics[scale=0.45]{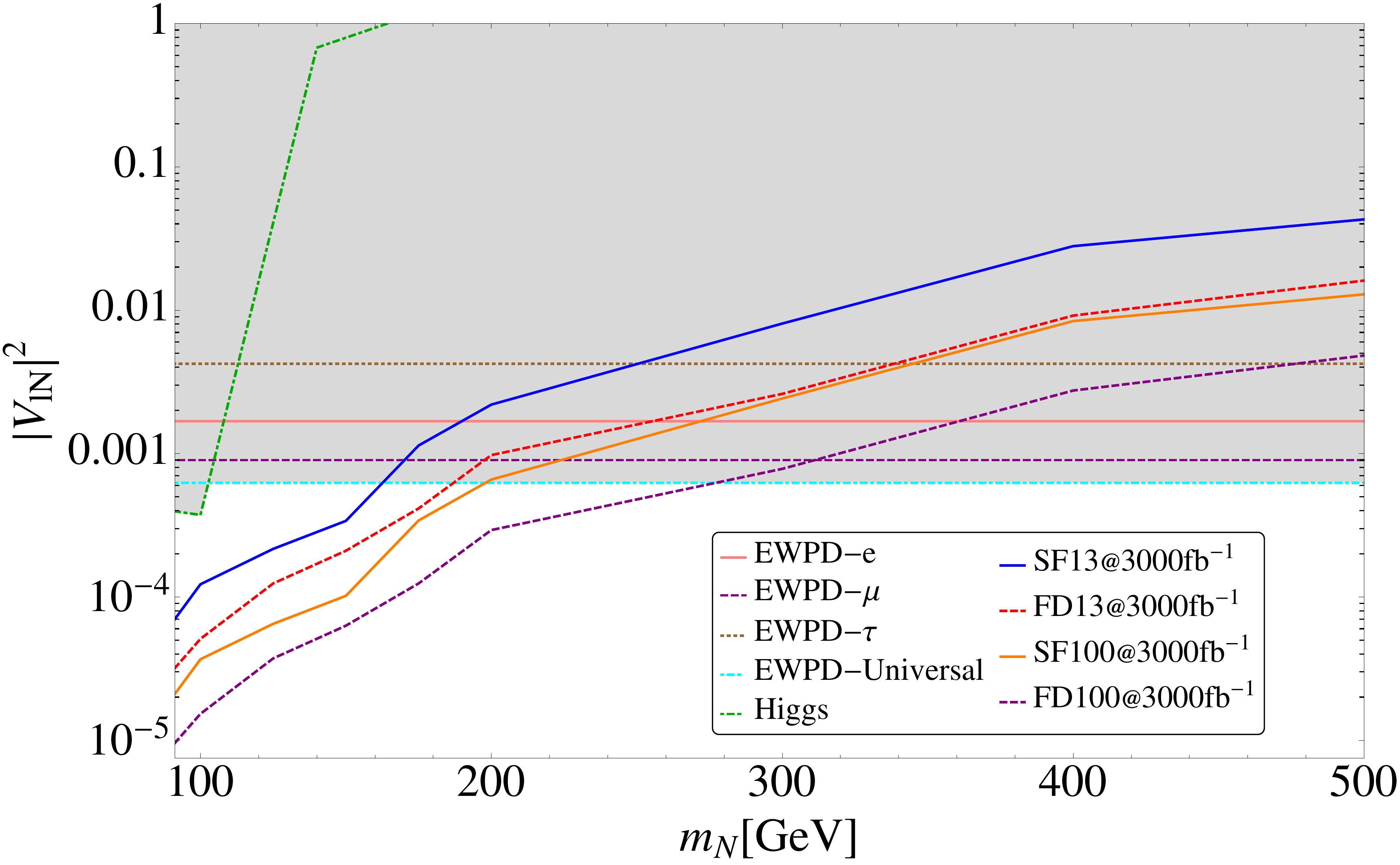}
\caption{Bounds on the square of the mixing angle from the trilepton plus MET final state as a function of the heavy neutrino mass $(m_N)$. The shaded region is ruled out by the EWPD and Higgs data. Prospective bounds from the Higgs
data had been calculated in \cite{Das:2017zjc} for the LHC at the high luminosity $(3000$ fb$^{-1})$ and future $100$ TeV pp collider. }
\label{mix1}
\end{figure}
 
The mixing-squared vs $m_N$ contours shown in Fig.~\ref{mix1} are allowed by the EWPD with in the range $91.2$ GeV$\leq m_N \leq 163$ GeV for the SF case at the 13 TeV LHC at 3000 fb$^{-1}$ luminosity. At the same time we analyze the FD case through which one can probe the range of $91.2$ GeV$\leq m_N \leq 195$ GeV. The limits for the FD case are roughly twice stronger than the SF case. However, at the future $100$ TeV pp collider with $3000$ fb$^{-1}$ luminosity the upper bounds on $m_{N}$ goes up to $200$ GeV and $275$ GeV for the SF and FD cases respectively keeping lower mass bound at the $91.2$ GeV. As the coupling between the Higgs and RHNs can not distinguish between the Majorana and pseudo-Dirac neutrinos, we use the same limits for both of the cases. Similarly the EWPD can measure the deviation from the SM couplings so we case use the same bounds in both of the cases.

\section{Conclusions}
\label{concl}
In this paper we have studied both of the type-I and inverse Seesaw models where SM singlet RHNs are involved. The RHNs involved in the type-I Seesaw mechanism are Majorana type where as those are present in the inverse Seesaw model are 
pseudo-Dirac in nature. We have studied the production mechanisms of the RHNs from various initial states at the 13 TeV LHC and future $100$ TeV pp collider considering the square of the mixing angle as $10^{-4}$ when the RHNs are produced in association with charged lepton, neutral lepton along with the jets. It has been shown due to the prompt decay of Higgs boson $(m_N < m_H)$ into the RHNs and SM light neutrino, the cross section dominates over all the remaining production modes. 
Such heavy neutrinos can dominantly decay into the $\ell W$ mode at the colliders. Depending upon the nature of the model we pick up two distinguishing signals from the type-I Seesaw and inverse Seesaw in the form of same sign di-lepton plus di-jet and trilepton plus MET respectively. 

Using the recent searches by the CMS and ATLAS for the type-I Seesaw with same-sign di-muon and di-jet final state we put a prospective upper bound on the $|V_{\mu N}|^2$ for the 13 TeV LHC and future $100$ TeV pp collider at $3000$ fb$^{-1}$ luminosity. 
Applying the cuts used in the anomalous multi-lepton search done by the LHC, we can put upper bounds on the square of the mixing angle between the SM light and of the degenerate RHNs. We consider the SF and FD cases where the limits in FD case is twice stronger than the SF case.

We have noticed that our currently given projected limits on the square of the mixing angles are better than those obtained from different experiments and even at the current stage of the LHC. We expect improvements in the status at the HL-LHC and future $100$ TeV pp collider when multi-lepton final states will be studied from the decays of the RHNs which will lead us to a more optimistic conclusion.
\section{Conflicts of Interest}
  The author declares that there is no conflict of interest regarding the publication of this paper.
\acknowledgments
The Author would like to thank E. J. Chun, P. S. Bhupal Dev, Y. Gao, T. Kamon, P. Konar, R. N. Mohapatra, N. Okada, A. Thalapillil, U. K. Yang
for useful discussions at different times. The author would like to thank  A. G. Hessler, A. Ibarra, E. Molinaro and S. Vogl for useful
information and discussions regarding  \cite{Hessler:2014ssa}. 


\begin{thebibliography}{999}
\bibitem{Neut1}
K.~Abe et.~al. [T2K Collaboration] Phys.~Rev.~Lett.~107, 041801 (2011).
\bibitem{Neut2}
P.~Adamson et al. [MINOS Collaboration], Phys.~Rev.~Lett.~107, 181802 (2011).
\bibitem{Neut3}
Y.~Abe et al. [DOUBLE-CHOOZ Collaboration], Phys.~Rev.~Lett.~108, 131801 (2012).
\bibitem{Neut4}
F.~P.~An et al. [DAYA-BAY Collaboration], Phys.~Rev.~Lett.~108, 171803 (2012).
\bibitem{Neut5}
J.~K.~Ahn et al. [RENO Collaboration], Phys.~Rev.~Lett.~108, 191802 (2012).    
\bibitem{Neut6}
C.~Patrignani {\it et al.} [Particle Data Group],
  ``Review of Particle Physics,''
  Chin.\ Phys.\ C {\bf 40}, no. 10, 100001 (2016).
  doi:10.1088/1674-1137/40/10/100001
\bibitem{Weinberg:1979sa} 
  S.~Weinberg,
  ``Baryon and Lepton Nonconserving Processes,''
  Phys.\ Rev.\ Lett.\  {\bf 43}, 1566 (1979).
  doi:10.1103/PhysRevLett.43.1566  
 \bibitem{Seesaw0} 
  P.~Minkowski,
  ``$\mu \to e\gamma$ at a Rate of One Out of $10^{9}$ Muon Decays?,''
  Phys.\ Lett.\  {\bf 67B}, 421 (1977).
  doi:10.1016/0370-2693(77)90435-X  
  \bibitem{Seesaw1}
T.~Yanagida, ``Horizontal Symmetry and Masses of Neutrinos,'' Prog.~Theor.~Phys.~{\bf 64}, 1103 (1980).
\bibitem{Seesaw2}
J.~Schechter and J.~W.~F.~Valle, ``Neutrino Masses in SU(2) $\otimes$ U(1) Theories,'' Phys.~Rev.~D {\bf 22}, 2227 (1980). 
\bibitem{Seesaw3}
T.~Yanagida, in Proceedings of the Work- shop on the Unified Theory and the Baryon Number in the Universe (O.~Sawada and A.~Sugamoto, eds.), KEK, Tsukuba, Japan, 1979, p.~95.
\bibitem{Seesaw4}
M.~Gell-Mann, P. Ramond, and R.~Slansky, Supergravity (P.~van Nieuwenhuizen et al.~eds.), North Holland, Amsterdam, 1979, p.~315.
\bibitem{Seesaw5}
S.~L.~Glashow, The future of elementary particle physics, in Proceedings of the 1979 Carg`ese Summer Institute on Quarks and Leptons (M. Levy et al. eds.), Plenum Press, New York, 1980, p.~687.
\bibitem{Seesaw6}
R.~N.~Mohapatra and G.~Senjanovic, ``Neutrino Mass and Spontaneous Parity Violation,'' Phys.~Rev.~Lett.~{\bf 44}, 912 (1980).    
 \bibitem{Casas:2001sr} 
  J.~A.~Casas and A.~Ibarra,
  ``Oscillating neutrinos and muon$ ---> e$, gamma,''
  Nucl.\ Phys.\ B {\bf 618}, 171 (2001)
  doi:10.1016/S0550-3213(01)00475-8
  [hep-ph/0103065].  
   \bibitem{Ibarra:2010xw} 
  A.~Ibarra, E.~Molinaro and S.~T.~Petcov,
  ``TeV Scale See-Saw Mechanisms of Neutrino Mass Generation, the Majorana Nature of the Heavy Singlet Neutrinos and $(\beta\beta)_{0\nu}$-Decay,''
  JHEP {\bf 1009}, 108 (2010)
  doi:10.1007/JHEP09(2010)108
  [arXiv:1007.2378 [hep-ph]].    
   
  
  \bibitem{Das:2017nvm} 
  A.~Das and N.~Okada,
  ``Bounds on heavy Majorana neutrinos in type-I Seesaw and implications for collider searches,''
  arXiv:1702.04668 [hep-ph]. 
   
  
  
  
  
  
  \bibitem{Adam:2011ch} 
  J.~Adam {\it et al.} [MEG Collaboration],
  ``New limit on the lepton-flavour violating decay $\mu^{+} \to e^{+} \gamma$,''
  Phys.\ Rev.\ Lett.\  {\bf 107}, 171801 (2011)
  doi:10.1103/PhysRevLett.107.171801
  [arXiv:1107.5547 [hep-ex]].  
  \bibitem{Aubert:2009ag} 
  B.~Aubert {\it et al.} [BaBar Collaboration],
  ``Searches for Lepton Flavor Violation in the Decays tau$+- --->$ e$+-$ gamma and tau$+- --->$ mu$+-$ gamma,''
  Phys.\ Rev.\ Lett.\  {\bf 104}, 021802 (2010)
  doi:10.1103/PhysRevLett.104.021802
  [arXiv:0908.2381 [hep-ex]].
  \bibitem{OLeary:2010hau} 
  B.~O'Leary {\it et al.} [SuperB Collaboration],
  ``SuperB Progress Reports -- Physics,''
  arXiv:1008.1541 [hep-ex].

  \bibitem{delAguila:2008pw} 
  F.~del Aguila, J.~de Blas and M.~Perez-Victoria,
  ``Effects of new leptons in Electroweak Precision Data,''
  Phys.\ Rev.\ D {\bf 78}, 013010 (2008)
  doi:10.1103/PhysRevD.78.013010
  [arXiv:0803.4008 [hep-ph]].  
   \bibitem{Akhmedov:2013hec} 
  E.~Akhmedov, A.~Kartavtsev, M.~Lindner, L.~Michaels and J.~Smirnov,
  ``Improving Electro-Weak Fits with TeV-scale Sterile Neutrinos,''
  JHEP {\bf 1305}, 081 (2013)
  doi:10.1007/JHEP05(2013)081
  [arXiv:1302.1872 [hep-ph]]. 
 \bibitem{Antusch:2006vwa} 
  S.~Antusch, C.~Biggio, E.~Fernandez-Martinez, M.~B.~Gavela and J.~Lopez-Pavon,
  ``Unitarity of the Leptonic Mixing Matrix,''
  JHEP {\bf 0610}, 084 (2006)
  doi:10.1088/1126-6708/2006/10/084
  [hep-ph/0607020]. 
 \bibitem{Abada:2007ux} 
  A.~Abada, C.~Biggio, F.~Bonnet, M.~B.~Gavela and T.~Hambye,
  ``Low energy effects of neutrino masses,''
  JHEP {\bf 0712}, 061 (2007)
  doi:10.1088/1126-6708/2007/12/061
  [arXiv:0707.4058 [hep-ph]].     
   \bibitem{Ibarra:2011xn} 
  A.~Ibarra, E.~Molinaro and S.~T.~Petcov,
  ``Low Energy Signatures of the TeV Scale See-Saw Mechanism,''
  Phys.\ Rev.\ D {\bf 84}, 013005 (2011)
  doi:10.1103/PhysRevD.84.013005
  [arXiv:1103.6217 [hep-ph]].  
  \bibitem{Dinh:2012bp} 
  D.~N.~Dinh, A.~Ibarra, E.~Molinaro and S.~T.~Petcov,
  ``The $\mu - e$ Conversion in Nuclei, $\mu \to e \gamma, \mu \to 3e$ Decays and TeV Scale See-Saw Scenarios of Neutrino Mass Generation,''
  JHEP {\bf 1208}, 125 (2012)
  Erratum: [JHEP {\bf 1309}, 023 (2013)]
  doi:10.1007/JHEP09(2013)023, 10.1007/JHEP08(2012)125
  [arXiv:1205.4671 [hep-ph]].  
   \bibitem{Penedo:2017knr} 
  J.~T.~Penedo, S.~T.~Petcov and T.~Yanagida,
  ``Low-Scale Seesaw and the CP Violation in Neutrino Oscillations,''
  arXiv:1712.09922 [hep-ph].  
\bibitem{Rasmussen:2016njh} 
  R.~W.~Rasmussen and W.~Winter,
  ``Perspectives for tests of neutrino mass generation at the GeV scale: Experimental reach versus theoretical predictions,''
  Phys.\ Rev.\ D {\bf 94}, no. 7, 073004 (2016)
  doi:10.1103/PhysRevD.94.073004
  [arXiv:1607.07880 [hep-ph]].

 
\bibitem{BhupalDev:2012zg} 
  P.~S.~Bhupal Dev, R.~Franceschini and R.~N.~Mohapatra,
  ``Bounds on TeV Seesaw Models from LHC Higgs Data,''
  Phys.\ Rev.\ D {\bf 86}, 093010 (2012)
  doi:10.1103/PhysRevD.86.093010
  [arXiv:1207.2756 [hep-ph]].
\bibitem{Aad:2015xaa} 
  G.~Aad {\it et al.} [ATLAS Collaboration],
  ``Search for heavy Majorana neutrinos with the ATLAS detector in pp collisions at $ \sqrt{s}=8 $ TeV,''
  JHEP {\bf 1507}, 162 (2015)
  doi:10.1007/JHEP07(2015)162
  [arXiv:1506.06020 [hep-ex]].
\bibitem{CMS8:2016olu}
V.~Khachatryan {\it et. al.} [CMS Collaboration],
``Search for heavy Majorana neutrinos in e$^{\pm}$e$^{\pm}$+jets and e$^{\pm}\mu^{\pm}$+jets events in proton-proton collisions at $\sqrt{s}=8$ TeV,"
JHEP {\bf 1604}, 169 (2016)
  doi: 10.1007/JHEP04 (2016)169
  [arxiv:1603.02248 [hep-ex]].
  
   
  \bibitem{Das:2017zjc} 
  A.~Das, P.~S.~B.~Dev and C.~S.~Kim,
  ``Constraining Sterile Neutrinos from Precision Higgs Data,''
  Phys.\ Rev.\ D {\bf 95}, no. 11, 115013 (2017)
  doi:10.1103/PhysRevD.95.115013
  [arXiv:1704.00880 [hep-ph]].  
  
    
  \bibitem{Ruiz:2017yyf} 
  R.~Ruiz, M.~Spannowsky and P.~Waite,
  ``Heavy neutrinos from gluon fusion,''
  Phys.\ Rev.\ D {\bf 96}, no. 5, 055042 (2017)
  doi:10.1103/PhysRevD.96.055042
  [arXiv:1706.02298 [hep-ph]].  
  
  

  \bibitem{Das:2017rsu} 
  A.~Das, Y.~Gao and T.~Kamon,
  ``Heavy Neutrino Search via the Higgs boson at the LHC,''
  arXiv:1704.00881 [hep-ph].  
  

  
  \bibitem{Kersten:2007vk} 
  J.~Kersten and A.~Y.~Smirnov,
  ``Right-Handed Neutrinos at CERN LHC and the Mechanism of Neutrino Mass Generation,''
  Phys.\ Rev.\ D {\bf 76}, 073005 (2007)
  doi:10.1103/PhysRevD.76.073005
  [arXiv:0705.3221 [hep-ph]].  
  
  
  \bibitem{Xing:2009in} 
  Z.~z.~Xing,
  ``Naturalness and Testability of TeV Seesaw Mechanisms,''
  Prog.\ Theor.\ Phys.\ Suppl.\  {\bf 180}, 112 (2009)
  doi:10.1143/PTPS.180.112
  [arXiv:0905.3903 [hep-ph]].  
  
  \bibitem{He:2009ua} 
  X.~G.~He, S.~Oh, J.~Tandean and C.~C.~Wen,
  ``Large Mixing of Light and Heavy Neutrinos in Seesaw Models and the LHC,''
  Phys.\ Rev.\ D {\bf 80}, 073012 (2009)
  doi:10.1103/PhysRevD.80.073012
  [arXiv:0907.1607 [hep-ph]].  
  
  
  \bibitem{Deppisch:2010fr} 
  F.~F.~Deppisch and A.~Pilaftsis,
  ``Lepton Flavour Violation and theta(13) in Minimal Resonant Leptogenesis,''
  Phys.\ Rev.\ D {\bf 83}, 076007 (2011)
  doi:10.1103/PhysRevD.83.076007
  [arXiv:1012.1834 [hep-ph]].  
  
  \bibitem{Dev:2013oxa} 
  C.~H.~Lee, P.~S.~Bhupal Dev and R.~N.~Mohapatra,
  ``Natural TeV-scale left-right Seesaw mechanism for neutrinos and experimental tests,''
  Phys.\ Rev.\ D {\bf 88}, no. 9, 093010 (2013)
  doi:10.1103/PhysRevD.88.093010
  [arXiv:1309.0774 [hep-ph]].  
  
  
  \bibitem{Mohapatra:1986aw} 
  R.~N.~Mohapatra,
  ``Mechanism for Understanding Small Neutrino Mass in Superstring Theories,''
  Phys.\ Rev.\ Lett.\  {\bf 56}, 561 (1986).
  doi:10.1103/PhysRevLett.56.561  
  
  
  \bibitem{Mohapatra:1986bd} 
  R.~N.~Mohapatra and J.~W.~F.~Valle,
  ``Neutrino Mass and Baryon Number Nonconservation in Superstring Models,''
  Phys.\ Rev.\ D {\bf 34}, 1642 (1986).
  doi:10.1103/PhysRevD.34.1642  
  
    
  \bibitem{Das:2012ze} 
  A.~Das and N.~Okada,
  ``Inverse Seesaw neutrino signatures at the LHC and ILC,''
  Phys.\ Rev.\ D {\bf 88}, 113001 (2013)
  doi:10.1103/PhysRevD.88.113001
  [arXiv:1207.3734 [hep-ph]].  
  
  \bibitem{Das:2014jxa} 
  A.~Das, P.~S.~Bhupal Dev and N.~Okada,
  ``Direct bounds on electroweak scale pseudo-Dirac neutrinos from $\sqrt s=8$ TeV LHC data,''
  Phys.\ Lett.\ B {\bf 735}, 364 (2014)
  doi:10.1016/j.physletb.2014.06.058
  [arXiv:1405.0177 [hep-ph]].  
  
  \bibitem{Das:2015toa} 
  A.~Das and N.~Okada,
  ``Improved bounds on the heavy neutrino productions at the LHC,''
  Phys.\ Rev.\ D {\bf 93}, no. 3, 033003 (2016)
  doi:10.1103/PhysRevD.93.033003
  [arXiv:1510.04790 [hep-ph]].  
  
  
  \bibitem{Das:2016akd} 
  A.~Das, N.~Nagata and N.~Okada,
  ``Testing the 2-TeV Resonance with Trileptons,''
  JHEP {\bf 1603}, 049 (2016)
  doi:10.1007/JHEP03(2016)049
  [arXiv:1601.05079 [hep-ph]].  
  
  
  \bibitem{Das:2016hof} 
  A.~Das, P.~Konar and S.~Majhi,
  ``Production of Heavy neutrino in next-to-leading order QCD at the LHC and beyond,''
  JHEP {\bf 1606}, 019 (2016)
  doi:10.1007/JHEP06(2016)019
  [arXiv:1604.00608 [hep-ph]].  
 
 
 \bibitem{Arkani-Hamed:2015vfh} 
  N.~Arkani-Hamed, T.~Han, M.~Mangano and L.~T.~Wang,
  ``Physics opportunities of a 100 TeV proton?proton collider,''
  Phys.\ Rept.\  {\bf 652}, 1 (2016)
  doi:10.1016/j.physrep.2016.07.004
  [arXiv:1511.06495 [hep-ph]]. 
 
  
  
    \bibitem{Catani:2001cc} 
  S.~Catani, F.~Krauss, R.~Kuhn and B.~R.~Webber,
  ``QCD matrix elements + parton showers,''
  JHEP {\bf 0111}, 063 (2001)
  doi:10.1088/1126-6708/2001/11/063
  [hep-ph/0109231]. 
 
  \bibitem{Das:2017pvt} 
  A.~Das,
  ``Pair production of heavy neutrinos in next-to-leading order QCD at the hadron colliders in the inverse Seesaw framework,''
  arXiv:1701.04946 [hep-ph].  
 
\bibitem{Matching}
J.~Alwal,~MadGraph Wiki, Matching of Jets between MadFevent and Pythia, 2011-10-05,~\url{https://cp3.irmp.ucl.ac.be/projects/madgraph/wiki/Matching}
\bibitem{Matching1}
J.~Alwall, S.~D.~Visscher,MadGraph Wiki, Introduction to jet-parton matching in MG/ME, 2011-03-16,~\url{https://cp3.irmp.ucl.ac.be/projects/madgraph/wiki/IntroMatching}
\bibitem{Matching2}
J.~Alwall, IPMU, Focus week, Tokyo, Japan, 12 November 2009,~\url{http://www.ipmu.jp/sites/default/files/webfm/pdfs/FocusWeek\_10}
\bibitem{Matching3}
J.~Alwall, Parton Showers and MLM matching with MadGraph and Pythia, NTU-MadGraph School, May 25-27, 2012.~\url{https://cp3.irmp.ucl.ac.be/projects/madgraph/attachment/.../NTU-MLM-lectures.pdf}
\bibitem{Matching31}
 J.~Alwall, HELAS/MadGraph Workshop, KEK 18-27 October 2006,~\url{http://phya.snu.ac.kr/~particle/wp-content/uploads/2007/03/lectureb-1.pdf}
\bibitem{Matching32}
J.~Alwall, HELAS/MadGraph Workshop, KEK 18-27 October 2006,~\url{http://phya.snu.ac.kr/~particle/wp-content/uploads/2007/03/lectureb-2.pdf}
\bibitem{Matching33}
J.~Alwall, HELAS/MadGraph Workshop, KEK 18-27 October 2006,~\url{http://phya.snu.ac.kr/~particle/wp-content/uploads/2007/03/lectureb-3.pdf}
\bibitem{Matching34}
M.~L.~Mangano, M.~Moretti, F.~Piccinini, R.~Pittau, ALPGEN, \url{http://mlm.web.cern.ch/mlm/talks/kek-alpgen.pdf}, 
~\url{http://mlm.home.cern.ch/mlm/alpgen}
\bibitem{Mangano:2002ea} 
  M.~L.~Mangano, M.~Moretti, F.~Piccinini, R.~Pittau and A.~D.~Polosa,
  ``ALPGEN, a generator for hard multiparton processes in hadronic collisions,''
  JHEP {\bf 0307}, 001 (2003)
  doi:10.1088/1126-6708/2003/07/001
  [hep-ph/0206293].

\bibitem{Mangano:2006rw} 
  M.~L.~Mangano, M.~Moretti, F.~Piccinini and M.~Treccani,
  ``Matching matrix elements and shower evolution for top-quark production in hadronic collisions,''
  JHEP {\bf 0701}, 013 (2007)
  doi:10.1088/1126-6708/2007/01/013
  [hep-ph/0611129].

\bibitem{Cooper:2011gk} 
  B.~Cooper, J.~Katzy, M.~L.~Mangano, A.~Messina, L.~Mijovic and P.~Skands,
  ``Importance of a consistent choice of alpha(s) in the matching of AlpGen and Pythia,''
  Eur.\ Phys.\ J.\ C {\bf 72}, 2078 (2012)
  doi:10.1140/epjc/s10052-012-2078-y
  [arXiv:1109.5295 [hep-ph]].

\bibitem{Alwall:2007fs} 
  J.~Alwall {\it et al.},
  ``Comparative study of various algorithms for the merging of parton showers and matrix elements in hadronic collisions,''
  Eur.\ Phys.\ J.\ C {\bf 53}, 473 (2008)
  doi:10.1140/epjc/s10052-007-0490-5
  [arXiv:0706.2569 [hep-ph]].
  

 \bibitem{Alwall:2011uj} 
  J.~Alwall, M.~Herquet, F.~Maltoni, O.~Mattelaer and T.~Stelzer,
  ``MadGraph 5 : Going Beyond,''
  JHEP {\bf 1106}, 128 (2011)
  doi:10.1007/JHEP06(2011)128
  [arXiv:1106.0522 [hep-ph]]. 
  
 \bibitem{Alwall:2008qv} 
  J.~Alwall, S.~de Visscher and F.~Maltoni,
  ``QCD radiation in the production of heavy colored particles at the LHC,''
  JHEP {\bf 0902}, 017 (2009)
  doi:10.1088/1126-6708/2009/02/017
  [arXiv:0810.5350 [hep-ph]].
  
  \bibitem{Matching4}
  \url{http://home.thep.lu.se/~torbjorn/pythia81html/JetMatching.html}
  \bibitem{Matching5}  
   \url{http://www.tifr.res.in/~indiacms/indiacms-meetings/march-2009/Matching$\%$20Parton$\%$\\
    20Showers$\%$20and$\%$20Matrix$\%$20elements.pdf}
    
\bibitem{Hoche:2006ph} 
  S.~Hoeche, F.~Krauss, N.~Lavesson, L.~Lonnblad, M.~Mangano, A.~Schalicke and S.~Schumann,
  ``Matching parton showers and matrix elements,''
  doi:10.5170/CERN-2005-014.288
  hep-ph/0602031.  
  
  
  
 
 \bibitem{Sjostrand:2006za} 
  T.~Sjostrand, S.~Mrenna and P.~Z.~Skands,
  ``PYTHIA 6.4 Physics and Manual,''
  JHEP {\bf 0605}, 026 (2006)
  doi:10.1088/1126-6708/2006/05/026
  [hep-ph/0603175]. 
  
  
   \bibitem{Hessler:2014ssa} 
  A.~G.~Hessler, A.~Ibarra, E.~Molinaro and S.~Vogl,
  ``Impact of the Higgs boson on the production of exotic particles at the LHC,''
  Phys.\ Rev.\ D {\bf 91}, no. 11, 115004 (2015)
  doi:10.1103/PhysRevD.91.115004
  [arXiv:1408.0983 [hep-ph]].  
  
  
  \bibitem{Williams:1934ad} 
  E.~J.~Williams,
  ``Nature of the high-energy particles of penetrating radiation and status of ionization and radiation formulae,''
  Phys.\ Rev.\  {\bf 45}, 729 (1934).
  doi:10.1103/PhysRev.45.729.  
  
  \bibitem{vonWeizsacker:1934nji} 
  C.~F.~von Weizsacker,
  ``Radiation emitted in collisions of very fast electrons,''
  Z.\ Phys.\  {\bf 88}, 612 (1934).
  doi:10.1007/BF01333110.
  
  \bibitem{Budnev:1974de} 
  V.~M.~Budnev, I.~F.~Ginzburg, G.~V.~Meledin and V.~G.~Serbo,
  ``The Two photon particle production mechanism. Physical problems. Applications. Equivalent photon approximation,''
  Phys.\ Rept.\  {\bf 15}, 181 (1975).
  doi:10.1016/0370-1573(75)90009-5  
  
  
  
  \bibitem{Dev:2013wba} 
  P.~S.~B.~Dev, A.~Pilaftsis and U.~k.~Yang,
  ``New Production Mechanism for Heavy Neutrinos at the LHC,''
  Phys.\ Rev.\ Lett.\  {\bf 112}, no. 8, 081801 (2014)
  doi:10.1103/PhysRevLett.112.081801
  [arXiv:1308.2209 [hep-ph]].  
  
  
  \bibitem{Alva:2014gxa} 
  D.~Alva, T.~Han and R.~Ruiz,
  ``Heavy Majorana neutrinos from $W\gamma$ fusion at hadron colliders,''
  JHEP {\bf 1502}, 072 (2015)
  doi:10.1007/JHEP02(2015)072
  [arXiv:1411.7305 [hep-ph]].
  
    \bibitem{Deppisch:2015qwa} 
  F.~F.~Deppisch, P.~S.~Bhupal Dev and A.~Pilaftsis,
  ``Neutrinos and Collider Physics,''
  New J.\ Phys.\  {\bf 17}, no. 7, 075019 (2015)
  doi:10.1088/1367-2630/17/7/075019
  [arXiv:1502.06541 [hep-ph]].
  

  
  \bibitem{CTEQ}
J.~Pumplin, D.~R.~Stump, J.~Huston, H.~L.~Lai, P.~M.~Nadolsky, W.~K.~Tung, J.~High Energy Phys.~0207 (2002) 012, arXiv: hep-ph/0201195.  
  
  
  \bibitem{deFavereau:2013fsa} 
  J.~de Favereau {\it et al.} [DELPHES 3 Collaboration],
  ``DELPHES 3, A modular framework for fast simulation of a generic collider experiment,''
  JHEP {\bf 1402}, 057 (2014)
  doi:10.1007/JHEP02(2014)057
  [arXiv:1307.6346 [hep-ex]].  
  

 

  
  \bibitem{Drewes:2013gca} 
  M.~Drewes,
  ``The Phenomenology of Right Handed Neutrinos,''
  Int.\ J.\ Mod.\ Phys.\ E {\bf 22}, 1330019 (2013)
  doi:10.1142/S0218301313300191
  [arXiv:1303.6912 [hep-ph]].  
  
   \bibitem{Dube:2017jgo} 
  S.~Dube, D.~Gadkari and A.~M.~Thalapillil,
  ``Lepton-Jets and Low-Mass Sterile Neutrinos at Hadron Colliders,''
  Phys.\ Rev.\ D {\bf 96}, no. 5, 055031 (2017)
  doi:10.1103/PhysRevD.96.055031
  [arXiv:1707.00008 [hep-ph]]. 
    
  \bibitem{Das:2017hmg} 
  A.~Das, P.~S.~B.~Dev and R.~N.~Mohapatra,
  ``Same Sign vs Opposite Sign Dileptons as a Probe of Low Scale Seesaw Mechanisms,''
  arXiv:1709.06553 [hep-ph].  
  
 \bibitem{Das:2017gke} 
  A.~Das, P.~Konar and A.~Thalapillil,
  ``Jet substructure shedding light on heavy Majorana neutrinos at the LHC,''
  arXiv:1709.09712 [hep-ph].
  
  \bibitem{Bhardwaj:2018lma} 
  A.~Bhardwaj, A.~Das, P.~Konar and A.~Thalapillil,
  ``Challenging Sterile Neutrino Searches at the LHC Complemented by Jet Substructure Techniques,''
  arXiv:1801.00797 [hep-ph].  
  
  
  \bibitem{Das:2017flq} 
  A.~Das, N.~Okada and D.~Raut,
  ``Enhanced pair production of heavy Majorana neutrinos at LHC,''
  arXiv:1710.03377 [hep-ph].    
  
 
  \bibitem{Das:2017deo} 
  A.~Das, N.~Okada and D.~Raut,
  ``Heavy Majorana neutrino pair productions at the LHC in minimal U(1) extended Standard Model,''
  arXiv:1711.09896 [hep-ph].  
  
  
    
   \bibitem{Mitra:2016kov} 
  M.~Mitra, R.~Ruiz, D.~J.~Scott and M.~Spannowsky,
  ``Neutrino Jets from High-Mass $W_R$ Gauge Bosons in TeV-Scale Left-Right Symmetric Models,''
  Phys.\ Rev.\ D {\bf 94}, no. 9, 095016 (2016)
  doi:10.1103/PhysRevD.94.095016
  [arXiv:1607.03504 [hep-ph]].   
  
  \bibitem{Das:2017kkm} 
  D.~Das, K.~Ghosh, M.~Mitra and S.~Mondal,
  ``Probing sterile neutrinos in the framework of inverse seesaw mechanism through leptoquark productions,''
  Phys.\ Rev.\ D {\bf 97}, no. 1, 015024 (2018)
  doi:10.1103/PhysRevD.97.015024
  [arXiv:1708.06206 [hep-ph]].  
  
  \bibitem{Drewes:2016jae} 
  M.~Drewes, B.~Garbrecht, D.~Gueter and J.~Klaric,
  ``Testing the low scale seesaw and leptogenesis,''
  JHEP {\bf 1708}, 018 (2017)
  doi:10.1007/JHEP08(2017)018
  [arXiv:1609.09069 [hep-ph]]. 
  
 \bibitem{Tang:2017plx} 
  Y.~L.~Tang,
  ``Probing the Light Sterile Neutrino Through the Heavy Charged Higgs Decay on the LHC,''
  arXiv:1712.10108 [hep-ph]. 
 
 
    
   
\bibitem{CMS8}
V.~Khachatryan {\it et al.} [CMS Collaboration], ``Search for heavy Majorana neutrinos in $\mu^\pm \mu^\pm$+jets events in proton-proton collisions at $\sqrt{s}$ = 8 TeV,''arXiv:1501.05566 [hep-ex].
\bibitem{CMS-trilep}
CMS Collaboration, arXiv:1404.5801 [hep-ex]; For an earlier analysis with $\sqrt{s} =$ 7 TeV LHC data, see S.~Chatrchyan, {\it et al.}, CMS
Collaboration, J.~High Energy Phys.~1206 (2012) 169, arXiv: 1204.5341 [hep-ex]. 

 \bibitem{deBlas:2013gla} 
  J.~de Blas,
  ``Electroweak limits on physics beyond the Standard Model,''
  EPJ Web Conf.\  {\bf 60}, 19008 (2013)
  doi:10.1051/epjconf/20136019008
  [arXiv:1307.6173 [hep-ph]]. 
  
  \bibitem{Adriani:1992pq} 
  O.~Adriani {\it et al.} [L3 Collaboration],
  ``Search for isosinglet neutral heavy leptons in Z0 decays,''
  Phys.\ Lett.\ B {\bf 295}, 371 (1992).
  doi:10.1016/0370-2693(92)91579-X  
  
  \bibitem{Acciarri:1999qj} 
  M.~Acciarri {\it et al.} [L3 Collaboration],
  ``Search for heavy isosinglet neutrinos in $e^{+} e^{-}$ annihilation at 130-GeV less than $S^{(1/2)}$ less than 189-GeV,''
  Phys.\ Lett.\ B {\bf 461}, 397 (1999)
  doi:10.1016/S0370-2693(99)00852-7
  [hep-ex/9909006].
  
  \bibitem{Achard:2001qv} 
  P.~Achard {\it et al.} [L3 Collaboration],
  ``Search for heavy isosinglet neutrino in $e^{+} e^{-}$ annihilation at LEP,''
  Phys.\ Lett.\ B {\bf 517}, 67 (2001)
  doi:10.1016/S0370-2693(01)00993-5
  [hep-ex/0107014].  
  
  \bibitem{Abreu:1996pa} 
  P.~Abreu {\it et al.} [DELPHI Collaboration],
  ``Search for neutral heavy leptons produced in Z decays,''
  Z.\ Phys.\ C {\bf 74}, 57 (1997)
  Erratum: [Z.\ Phys.\ C {\bf 75}, 580 (1997)].
  doi:10.1007/s002880050370  
  
  
  \bibitem{Blondel:2014bra} 
  A.~Blondel {\it et al.} [FCC-ee study Team],
  ``Search for Heavy Right Handed Neutrinos at the FCC-ee,''
  Nucl.\ Part.\ Phys.\ Proc.\  {\bf 273-275}, 1883 (2016)
  doi:10.1016/j.nuclphysbps.2015.09.304
  [arXiv:1411.5230 [hep-ex]].  
  
  
  \bibitem{delAguila:2008pw} 
  F.~del Aguila, J.~de Blas and M.~Perez-Victoria,
  ``Effects of new leptons in Electroweak Precision Data,''
  Phys.\ Rev.\ D {\bf 78}, 013010 (2008)
  doi:10.1103/PhysRevD.78.013010
  [arXiv:0803.4008 [hep-ph]].
  
  
  \bibitem{Akhmedov:2013hec} 
  E.~Akhmedov, A.~Kartavtsev, M.~Lindner, L.~Michaels and J.~Smirnov,
  ``Improving Electro-Weak Fits with TeV-scale Sterile Neutrinos,''
  JHEP {\bf 1305}, 081 (2013)
  doi:10.1007/JHEP05(2013)081
  [arXiv:1302.1872 [hep-ph]].  
  
    
  
  \bibitem{Chatrchyan:2012ty} 
  S.~Chatrchyan {\it et al.} [CMS Collaboration],
  ``Search for the standard model Higgs boson decaying to $W^+W^-$ in the fully leptonic final state in pp collisions at $\sqrt{s}=7$ TeV,''
  Phys.\ Lett.\ B {\bf 710}, 91 (2012)
  doi:10.1016/j.physletb.2012.02.076
  [arXiv:1202.1489 [hep-ex]].
  
  
  \bibitem{CMS:2012xwa} 
  CMS Collaboration [CMS Collaboration],
  ``Search for the standard model Higgs boson decaying to a W pair in the
fully leptonic final state in pp collisions at $\sqrt(s) = 8$ TeV,''
  CMS-PAS-HIG-12-017.
  
  
  
  \bibitem{Aad:2012uub}
  G.~Aad {\it et al.} [ATLAS Collaboration],
  ``Search for the Standard Model Higgs boson in the $H \to$ WW(*) $\to \ell \nu \ell \nu$ decay mode with 4.7 /fb of ATLAS data at $\sqrt{s}=7$ TeV,''
  Phys.\ Lett.\ B {\bf 716} (2012) 62
  doi:10.1016/j.physletb.2012.08.010
  [arXiv:1206.0756 [hep-ex]].
  
  
  \bibitem{Chatrchyan:2012ft} 
  S.~Chatrchyan {\it et al.} [CMS Collaboration],
  ``Search for the standard model Higgs boson in the $H$ to $Z Z$ to 2 $\ell 2 \nu$ channel in $pp$ collisions at $\sqrt{s}=7$ TeV,''
  JHEP {\bf 1203}, 040 (2012)
  doi:10.1007/JHEP03(2012)040
  [arXiv:1202.3478 [hep-ex]].
  
  
  \bibitem{Aad:2012ora} 
  G.~Aad {\it et al.} [ATLAS Collaboration],
  ``Search for a standard sodel Higgs boson in the H $\to$ ZZ $\to \ell^{+}\ell^{-}\nu \bar{\nu}$ decay channel using 4.7 fb$^{-1}$ of $\sqrt{s} =$ 7 TeV data with the ATLAS detector,''
  Phys.\ Lett.\ B {\bf 717}, 29 (2012)
  doi:10.1016/j.physletb.2012.09.016
  [arXiv:1205.6744 [hep-ex]].  
      
    
    
    
    \end{thebibliography}
  \end{document}